\documentclass[reprint,amsmath,amssymb,amsfonts,aps,prx,floatfix,]{revtex4-2}
\usepackage{times}
\usepackage{graphicx}
\usepackage{dcolumn}
\usepackage{bm}
\usepackage[colorlinks=true, citecolor=blue, linkcolor=blue, urlcolor=blue]{hyperref}
\usepackage{xcolor}
\definecolor{purple}{rgb}{0.6, 0, 0.6}

\begin{document}

\title{Electron Pairing Induced by Repulsive Interactions in Tunable One-Dimensional Platforms}

\author{Gal Shavit}
\author{Yuval Oreg}
\affiliation{
 Department of Condensed Matter Physics, Weizmann Institute of Science, Rehovot, Israel 76100
}

\date{\today}

\begin{abstract}
We present a scheme comprised of a one-dimensional system with repulsive interactions, in which the formation of bound pairs can take place in an easily tunable fashion.
By capacitively coupling a primary electronic quantum wire of interest to a secondary strongly-correlated fermionic system, the intrinsic electron-electron repulsion may be overcome, promoting the formation of bound electron pairs in the primary wire.
The intrinsic repulsive interactions tend to favor the formation of charge density waves of these pairs, yet we find that superconducting correlations are dominant in a limited parameter regime.
Our analysis show that the paired phase is stabilized in an intermediate region of phase space, encompassed by two additional phases: a decoupled phase, where the primary wire remains gapless, and a trion phase, where a primary electron pair binds a charge carrier from the secondary system.
Tuning the strength of the primary-secondary interaction, as well as the chemical potential of the secondary system, one can control the different phase transitions.
Our approach takes into account the interactions among the secondary degrees of freedom, and strongly relies on their highly correlated nature.
Extension of our proposal to two dimensions is discussed, and the conditions for a long-range superconducting order from repulsion only are found.
Our physical description, given by a simple model with a minimal amount of ingredients, may help to shed some light on pairing mechanism in various low-dimensional strongly correlated materials.

\end{abstract}

\maketitle

\section{\label{sec:intro}Introduction\protect}

Ever since Bardeen, Cooper, and Schrieffer (BCS) put forward their theory of superconductivity \cite{BCSref}, the means by which lattice electrons may form pairs despite their bare Coulomb repulsion~\cite{Tolmachev1961,Morel1962} has been an issue of significant interest in condensed matter physics.
Whereas conventional superconductors are fairly well described by BCS theory, with lattice phonons mediating the retarded electron-electron attractive interactions, high $T_c$ superconductors are most likely the result of a different electron pairing mechanisms.
These are usually related to the strongly correlated nature of the system charge carriers~\cite{HighTcpairingbook}, and are possibly related to strong spin fluctuations~\cite{highTcAFfluctuations}, or to coupling of the carriers to degrees of freedom which are in close vicinity to a quantum critical point~\cite{highTcQCPspinfermion,HighTcnematicQCP}.

The concept of short range attraction mediated by purely electronic degrees of freedom has been around for some time~\cite{LittleWA, PairingGinzburg, PairingGinzburgExcitonTcisHigher,PairingBardeenExciton1973,PairingBeni1974,PairingIonova1977, PairingLittle1981,PairingLittleGutfreund1996, VayrynenGoldsteinGefen}.
The pioneering work by Little~\cite{LittleWA} discussed the possibility of an organic superconductor, where pairing occurred due to the coupling of electrons in a one-dimensional (1D) molecule to polarizable side-chains.
This idea was implemented in a recent experiment~\cite{Hamo2016}, where a sizable pairing gap in a two-site system was indeed induced by Coulomb interactions with a nearby two-level quantum dot.
Whether or not this effective attractive interaction gives rise to superconductivity, especially in the context of 1D systems (where long ranged order superconductivity is inevitably absent~\cite{MerminWagnerHoenberg}) is contemplated in the literature~\cite{HirschScalpainoPRL}.

The importance of this concept is two-fold. First, since electron-electron repulsion is responsible for the pairing mechanism, the size of the gap can potentially be much higher as compared to conventional electron-phonon superconductors.
Hopefully, the magnitude of the Coulomb interactions in low-dimensional strongly correlated materials may even be large enough as to plausibly enable engineering superconductivity at room temperature and ambient pressure.
Secondly, as the condensed matter community still investigates the role of strong correlation effects on the origin of high-$T_c$ superconductors, as well as their phase diagram, models with so-called ``excitonic'' pairing, as we study here, may shed some light on such materials, and lead to new physical insights.

\begin{figure}
\includegraphics[scale=0.33]{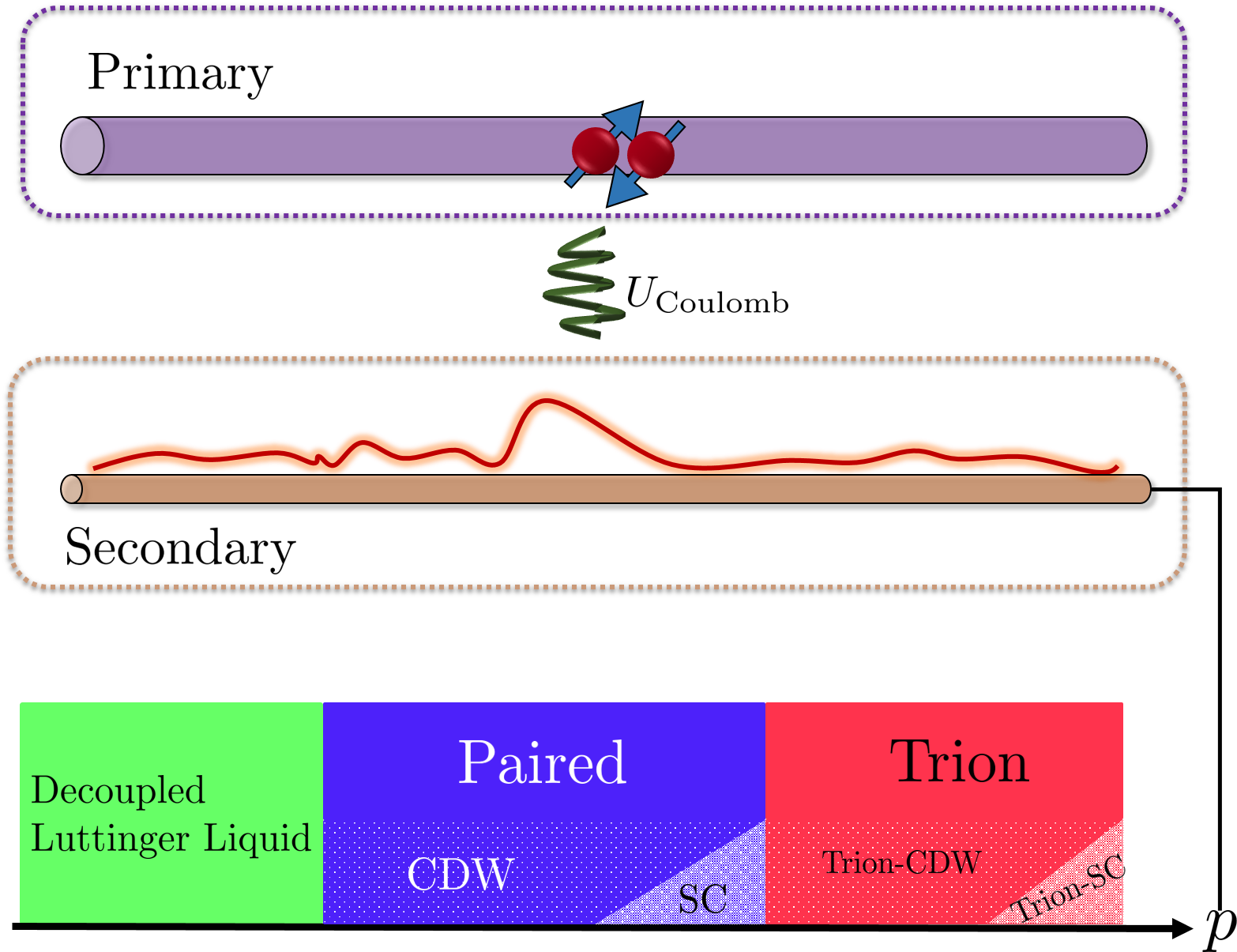}
\caption{\label{fig:MainSchematic}
Schematic description of our proposed setup. The spinful electrons in the 1D primary system (purple) interact via the Coulomb force with an itinerant many-body excitation (indicated by the orange charge distribution) in a strongly correlated secondary wire (brown), such as a spin less 1D system with strong repulsive interactions .
By carefully tuning the secondary wire, using experimentally accessible knobs, e.g., gates and magnetic fields, one can control the phase of the system and tune it between the decoupled Luttinger liquid phase (where the primary wire is gapless), the paired phase (with primary spin-singlet electron pairs),   and the trion phase (where a primary electron pair is bound to a secondary quasi-particle).
This is indicated in the bottom panel, where as a function of some tunable parameter $p$ the phase of the system is modified.
Whether the charge-density wave (CDW) or superconducting (SC) tendencies prevail within the paired and trion phases typically depends on intrinsic non-tunable parameters, e.g., the strength of interactions within the primary wire.
}
\end{figure}

In this work, we present (Sec.~\ref{sec:model}) a model for a  ``primary'' system of a spinful 1D wire with strong repulsive electron-electron interactions, in which pairing can be externally induced and readily controlled. Our RG analysis (Sec.~\ref{sec:RGanalysis}) shows that the existence of the electron pairs heavily relies on coupling of the primary system to a strongly correlated ``secondary'' system, which possesses a high degree of tunability, see Fig. \ref{fig:MainSchematic}. More concretely, we propose that a 1D system with strongly interacting spinless (or spin-polarized) electrons (e.g., a suspended magnetic-field-tuned carbon nanotube \cite{EggerCnt}) is suitable to play this secondary role. We use the RG approach to characterize the different interaction strengths and Fermi momenta incommensurabilities which determine the fate of the overall system.
The strong correlations between the secondary degrees of freedom themselves become essential in our scheme, and allows one to manipulate the phase of the primary system of interest with much ease.
The proposed setup can be realized in available 1D experimental platforms, e.g., carbon nanotubes \cite{CNTIijima1991}.

In contrast to previous works, which studied setups involving the engineering of coupled interacting nanowires for the purpose of implementing topological superconductivity or parafermions~\cite{MongFisher11_2014,KesselmanHaimBerg2014,Flensberg2014,Flensberg20142,Loss2014,HelicalSelaOreg,KesselmanBerg2015,MengThomale2015,Vekua2016,Sagi2017,RuhmanFu2017,TritopsReview}, our model relies solely on (i) \textit{capacitive} coupling between the wires, i.e., no tunneling of electrons between them, and (ii) strong purely repulsive electron-electron interactions.

We find that our theoretical model may lead to three distinct phases of the overall system, which we discuss in Sec.~\ref{sec:phasediagram}. One, almost trivial, is the ``decoupled'' phase, in which the primary and secondary systems are decoupled and the primary system is gapless. It is the consequence of either weak primary-secondary interactions, or a large enough mismatch of the two parts' Fermi momenta. In the opposite extreme, strong interactions between the primary and secondary systems and commensurability result in a ``locked'' trion phase, where a primary electron pair is bound to a secondary quasi-particle. This is manifested by opening of a partial gap in the total system, akin to those in Refs.~\cite{HelicalSelaOreg,SelaShotNoise,fractionalGSYO}.
In between the decoupled and trion phases an  ``electron-paired'' phase is established  by detuning the Fermi momenta. The detuning effectively breaks the composite particles of the trion phase, while keeping the primary electron pair intact. Depending on the interactions details the trion phase and the electron-paired phase can be dominated by charge density wave or superconducting correlations.
We would like to emphasize that, surprisingly, we find that strong interactions among the electrons in the secondary system enlarge the regions of both the trion and electron pair phases.
Unique experimental signatures for each phase, including single-particle tunneling gaps and fractional two-terminal electrical conductance, are also discussed.

In addition, we consider two important extensions of our model.
In Sec.~\ref{sec:Littlemodel}, we explore the relation between this work and the side-chains polarizers model proposed by Little.
In a certain extreme limit of our setup, namely a gapped non-itinerant secondary system, a clear connection can indeed be made. However, we emphasize that the high degree of tunability, the mobile nature of the secondary quasi-particles, and the role of strong repulsive interactions in the secondary system are essential ingredients which are accounted for in the model considered in this work.
In Sec.~\ref{sec:twoD}, by employing a coupled-wires approach, we elucidate under what circumstances an anisotropic two-dimensional variant of our setup may lead to superconductivity.
Conceptually, this bears some resemblance to the discussion of inter-stripe coherence in the cuprates \cite{Smectic}, with the main role of our work being the introduction of a novel mechanism by which the primary electrons form bound pairs within each 1D sector.

We conclude this work in Sec. \ref{sec:conclusions} and discuss the main implication of this work: experimentally feasible, highly tunable electron pairing in 1D 'primary' systems can be realized by Coulomb interactions with strongly correlated fermions hosted in an externally tunable 'secondary' system.

\section{\label{sec:mainsec}Pairing From Repulsion \protect}
The pairing mechanism we propose in this work is of extrinsic origin. It is the consequence of capacitively coupling a primary system of interest, e.g., a quantum nanowire or a 1D constriction in a two-dimensional electron gas, to a tunable strongly interacting secondary platform, such as a carbon nanotube~\cite{Bockrath1999CNTluttingerliquid}.
Modifying some properties of the secondary portion, one is able to enhance or suppress pairing and to manipulate the overall phase diagram. In this Section, we give a detailed analysis of such a minimal setup giving rise to controlled, tunable electron pairing. The various phases of the overall system, the decoupled, electron-paired, and trion phases, are explored, and their properties are analyzed.

\subsection{\label{sec:model} Minimal model \protect}

The setup we suggest is comprised of  three important ingredients: (i) A spinful
interacting semiconducting quantum wire, or any other 1D system with two electron species; (ii) A secondary one-dimensional
system which is either gapless, or has an experimentally controllable gap;
(iii) Interactions between electrons in the primary wire and the
(quasi-) particles of the secondary part.

Overall, the setup we propose is described by a Hamiltonian of the form
\begin{equation}
    H = \int dx \left[\mathcal{H}_{\rm pri} +\mathcal{H}_{\rm sec}+\mathcal{H}_{\rm int}\right]. \label{eq:Htotal}
\end{equation}
The primary wire degrees of freedom are described by
\begin{equation}
    \mathcal{H}_{\rm pri}=\mathcal{H}^0_{\rm pri}+\mathcal{H}^g_{\rm pri}.\label{eq:Hsys}
\end{equation}
$\mathcal{H}^0_{\rm pri}$, the non-interacting part, is given by the continuum description
\begin{align}
        \mathcal{H}^0_{\rm pri} & = iv_F\sum_{\sigma}\left(\psi_{R\sigma}^\dagger\partial_{x} \psi_{R\sigma}-\psi_{L\sigma}^\dagger \partial_x \psi_{L\sigma}\right)\nonumber\\
        & - \mu \sum_{\sigma}\left(\psi_{R\sigma}^\dagger \psi_{R\sigma}+\psi_{L\sigma}^\dagger\psi_{L\sigma}\right)\label{eq:Hsys0},
\end{align}
where $\psi_{R/L\sigma}=\psi_{R/L\sigma}\left(x\right)$ annihilates a right/left-moving electron with spin $\sigma$ at position $x$. The parameters, $v_F,\mu $ are the Fermi velocity and chemical potential, respectively. Interactions can be accounted for in the general form
\begin{equation}
    \mathcal{H}^g_{\rm pri}=g^{\alpha\beta}_{\gamma\delta}\psi^\dagger_\alpha\psi^\dagger_\beta\psi_\gamma\psi_\delta,\label{eq:Hsysg},
\end{equation}
 with greek indices generalizing chirality and spin, and repeated indices are summed over. The elements of the interaction matrix $g^{\alpha\beta}_{\gamma\delta}$ determines the nature of interactions in the system, and we assumed that the system is away from half-filling so we do not include umpklapp processes. Notice that for simplicity we have also implicitly assumed the existence of short-range interactions only, this will not have an important effect on our results.

The Hamiltonian term $\mathcal{H}_{\rm sec}$ describes the physics of the secondary system. In this work, we consider a 1D system hosting strongly interacting spinless fermions,
with chiral annihilation operators $c_{R/L}\left( x \right)$ at position $x$,
\begin{align}
        \mathcal{H}_{\rm sec} & = iv_{\rm sec}\left(c_{R}^\dagger\partial_{x} c_{R}-c_{L}^\dagger \partial_x c_{L}\right)\nonumber\\
        & - \mu_{\rm sec} \rho_{\rm sec}+m_{\rm sec}\left(c_R^\dagger c_L+\rm{h.c.}\right)\nonumber\\
        & +\int dx' \rho_{\rm sec} \left(x\right) U\left(\left| x-x'\right|\right)\rho_{\rm sec} \left(x'\right) \label{eq:Haux},
\end{align}
where $v_{\rm sec}$, $\mu_{\rm sec}$, $m_{\rm sec}$ are the secondary Fermi velocity, chemical potential, and mass, respectively. We have explicitly allowed here for finite-range interactions, (so that the corresponding  Luttinger liquid parameter that we introduce later can be smaller than 1/2) , with the charge density in the secondary wire $\rho_{\rm sec}\equiv c_{R}^\dagger c_{R} +c_{L}^\dagger c_{L}$, and the interaction strength determined by the function $U$. An experimentally accessible platform to implement such a strongly-correlated effectively spinless 1D system is a carbon nanotube (CNT), tuned by gates and magnetic flux. Details for such an implementation are given in Appendix~\ref{app:CNTsecondaryproposal}.

We now include the most crucial part in our model, capacitive coupling between the primary electrons and the secondary fermions,
\begin{equation}
    \mathcal{H}_{\rm int}=\rho_{\rm sec} V \left(\rho_\uparrow+\rho_\downarrow\right),\label{eq:Hint}
\end{equation}
with  $V>0$ the interaction strength, and $\rho_\sigma\equiv \psi_{R\sigma}^\dagger \psi_{R\sigma}+\psi_{L\sigma}^\dagger\psi_{L\sigma}$. Again, we assume short-range interactions, which we find {\it a posteriori} to be sufficient in order to capture the most important consequences of this term and the way it determines the phases of the system.

The model Eq. \eqref{eq:Htotal} is best treated in the framework of abelian bosonization \cite{LLHaldane,1dFL,giamarchi2004quantum}. This is achieved by expressing the chiral fermionic operators in terms of bosonic variables,
\begin{equation}
    \psi_{r\sigma}=\frac{\eta_{r \sigma}}{2\pi\alpha}e^{i\left(\theta_\sigma-r\phi_\sigma+rk_{F}x\right)},\,\,\,\,\,   c_{r}=\frac{\tilde{\eta}_{r }}{2\pi\alpha}e^{i\left(\theta-r\phi+rkx\right)},\label{eq:bosonization}
\end{equation}
with $r=\pm$ corresponding to $R/L$, $\alpha$ is the short-distance cutoff of our continuum model, $k_F,k$ are the Fermi momenta of the electrons (with spin up and down) in the primary system and the fermions in the  secondary system, respectively, $\eta,\tilde{\eta}$ are Klein factors ensuring fermionic commutation relations, and the bosonic fields obey the algebra
\begin{equation}
    \left[\phi_i\left(x\right),\partial_x \theta_j\left(x'\right)\right]=i\pi\delta\left(x-x'\right)\delta_{i,j},\label{eq:bosoncommute}
\end{equation}
with the indices $i,j$ specifying one of the two primary $\uparrow/\downarrow$ sectors or the secondary sector.
In terms of bosons, we may write $\mathcal{H}_{\rm{pri}}  =\mathcal{H}_c+\mathcal{H}_s$, accounting for the charge and spin part ($c,s$) of the Hamiltonian,
\begin{equation}
    \mathcal{H}_c=\frac{u_c}{2\pi}\left[ \frac{1}{K_c}\left(\partial_x \phi_c\right)^2+K_c \left(\partial_x \theta_c\right)^2 \right],\label{eq:Hc}
\end{equation}
\begin{align}
        \mathcal{H}_s&=\frac{u_s}{2\pi}\left[ \frac{1}{K_s}\left(\partial_x \phi_s\right)^2+K_s \left(\partial_x \theta_s\right)^2 \right]\nonumber\\
        &+\frac{g_s}{2\pi^2\alpha^2}\cos\left(\sqrt{8}\phi_s\right), \label{eq:Hs}
\end{align}
where the charge and spin sectors were defined by $\phi_{c/s}=\frac{\phi_\uparrow\pm\phi_\downarrow}{\sqrt{2}}$ (and similarly for $\theta_{c/s}$). Eqs. \eqref{eq:Hc}--\eqref{eq:Hs} feature the famous spin-charge separation. For repulsive electron-electron interactions one usually finds~\footnote{often it is assumed $K_s=1$, $g_s=0$, which is the fixed point of the RG flow for a system with repulsive interactions and SU(2) symmetry \cite{giamarchi2004quantum}.} $K_c<1$, $K_s>1$, and $g_s>0$.  For the secondary sector we have  \begin{align}
        \mathcal{H}_{\rm sec}&=\frac{u}{2\pi}\left[ \frac{1}{K}\left(\partial_x \phi\right)^2+K \left(\partial_x \theta\right)^2 \right]\nonumber\\
        &+\frac{m_{\rm sec}}{2\pi\alpha}\cos\left(2\phi-2kx\right), \label{eq:Habososn}
\end{align}
where $u,K$ are determined by $v_{\rm sec}$ and $U$. In this work we will mostly assume strong (or sufficiently long-range) repulsive $U$, such that $K$ may be much smaller than 1. If $m_{\rm sec}\neq0$, changing $\mu_{\rm sec}$ (and as a consequence $k$) leads to a commensurate-incommensurate transition, where the gap induced by the mass term is opened or closed.

Finally, the potentially relevant parts of \eqref{eq:Hint} may be written in the general form
\begin{align}
    \mathcal{H}_{\rm int} &= \frac{V_\phi}{\pi^2}\partial_x \phi_c \partial_x \phi + \frac{V_\theta}{\pi^2}\partial_x \theta_c \partial_x \theta\nonumber\\
    & + \frac{g_1}{2\pi^2\alpha^2}\cos \left( \sqrt{2}\phi_s \right) \cos \left( \sqrt{2}\phi_c + 2\phi -2k_+x\right)\nonumber\\
    & + \frac{g_2}{2\pi^2\alpha^2}\cos \left( \sqrt{2}\phi_s \right) \cos \left( \sqrt{2}\phi_c - 2\phi -2k_- x\right),\label{eq:Hintboson}
\end{align}
with $k_\pm=k_F\pm k$. The first line of Eq.~\eqref{eq:Hintboson} is the so-called forward scattering interactions, whereas the last two terms involve the momentum transfer of $2k_F$ and $2k$. We note that for the kind of interaction shown in Eq.~\eqref{eq:Hint}, which is of a pure density-density type, one finds $V_\phi=\sqrt{2}V$, $V_\theta=0$, $g_1=g_2=\frac{V}{2}$.

We proceed with a short qualitative discussion of our model, Eqs.~\eqref{eq:Htotal},\eqref{eq:Hc}--\eqref{eq:Hintboson} and the possible phases that may be obtained in limiting cases.

If $k_F$ and $k$ are grossly mismatched, such that $k_\pm$ times a typical correlation length are much larger than 1, or if the secondary sector has a gap, the $g_1,g_2$ terms are irrelevant as they oscillate rapidly in space. Then, the spin sector decouples from the other two, and for generic repulsive interactions it will also be gapless ($g_s$ is marginally irrelevant). The remaining bilinear interaction terms between the charge of the primary wire and secondary degrees of freedom cannot open any gaps, yet they cause some mixing of these sectors, changing the relevant Luttinger liquid (LL) susceptibilities power-laws (see, e.g., Ref. \cite{LLmix1}).
We refer to this phase as the decoupled LL phase.

What happens when either $g_1$ or $g_2$ are relevant? From a semi-classical point of view, one sees that since  $g_s>0$, the $g_{1,2}$ term competes with the $g_s$ one. The minima of $\cos\left(\sqrt{8}\phi_s\right)$ at $\sqrt{8} \phi_s= \pi(2n+1)$, with $n$ an integer, lead to $\cos\left(\sqrt{2}\phi_s\right)=0$. Thus, there is no simple $\phi_s$ configuration which minimizes the classical energy of both such terms.
As will be shown, this is a signature of the competition between the pairing tendency in the primary system and the intrinsic electron-electron repulsion within it.
In the limit $g_{1/2}\to\infty$ the terms in the corresponding cosines are pinned. The spin sector is gapped, as well as an additional sector combining the charge and secondary degrees of freedom. More concretely, if it is $g_1$ that is at the strong coupling fixed point, the only remaining gapless sector is $\propto \phi_\uparrow +\phi_\downarrow -\phi$. We refer to this phase as the ``trion'' phase, since the gapless sector may be interpreted as a Luttinger liquid of composite particles comprised of a pair of electrons bound to a secondary hole. This is to be contrasted with bound electrons composite particles induced by attraction in other 1D platforms~\cite{Briggeman2020}.

In our analysis below we will establish an additional intermediate phase, where electrons with opposite spins indeed form bound pairs, yet $g_{1,2}$ are irrelevant at low enough energy scales, leaving the secondary sector essentially unaltered. The spin-gapped ``paired'' phase that develops in the primary wire is reminiscent of the Luther-Emery liquid~\cite{LutherEmeryLiquid}, which is generic in spinful systems with attractive interactions, e.g., in the negative-$U$ 1D Hubbard model, however, here it is established for repulsive interaction only.

\subsection{\label{sec:RGanalysis} RG analysis \protect}
It turns out to be useful to introduce a unitary transformation to our Hamiltonian $H=\mathcal{H}_{\rm pri}+\mathcal{H}_{\rm sec}+\mathcal{H}_{\rm int}$ in their bosonized form. To do so consider
\[
\tilde{U}_Q={\rm exp}\left[{-iQ\int dx \partial_x\theta_c \phi}\right],
\]
where $Q$ parameterizes the transformation. Applying this transformation to the Hamiltonian $\tilde{U}_Q^\dagger H \tilde{U}_Q$ has the effect (c.f. Ref. \cite{SpinGapProximityEffect})
\[
\phi_c\to\phi_c-Q\phi,\,\,\,\,\,\partial_x \theta\to\partial_x \theta+Q\partial_x \theta_c,
\]
and modifies the Hamiltonian accordingly. We now specialize our theory to a specific form of the repulsive interactions between the primary and the secondary system and assume
\begin{equation}
    V_\theta=-V_\phi \frac{u}{u_{c}}K_{c}K\label{eq:specialV}.
\end{equation}
Along this line in the $V_\phi-V_\theta$ plane, we choose $Q=\frac{V_\phi K_{c}}{\pi u_{c}}$, allowing us to completely eliminate the forward scattering part of  $\mathcal{H}_{\rm int}$. This will simplify our analysis significantly, and enable a non-perturbative treatment of $ V_\phi$.
Generic repulsive interactions that somewhat deviate from Eq.~\eqref{eq:specialV} will influence the phase diagram only quantitatively and not qualitatively.
Defining the dimensionless interaction parameter $\upsilon\equiv\frac{K_c K}{\pi \sqrt{u_c u}} V_\phi$, we can sum up the changes to the Hamiltonian as \begin{subequations}
\begin{equation}
    K_c \to \tilde{K}_c=\sqrt{1-\left(\frac{u}{u_c}\right)^2\upsilon^2}K_c,\label{eq:KcTransformed}
\end{equation}
\begin{equation}
    K \to \tilde{K}=\sqrt{\frac{1}{1-\upsilon^2}}K,
\end{equation}
\begin{equation}
    \cos \left( \sqrt{2}\phi_c \pm 2\phi \right)\to \cos \left( \sqrt{2}\phi_c \pm 2\left(1\mp\frac{Q}{\sqrt{2}}\right)\phi\right),
\end{equation}
\end{subequations}
where in the last cosine terms the $k_\pm$ dependence is implicit.

We analyze our model using perturbative renormalization group (RG), up to second order in all couplings. Before deriving the full RG equations, we should address the issue of the oscillating cosines which appear throughout the theory. A fair approximation is to treat the momenta $k_\pm$ (and $k$ for the mass term, if it exists) as the inverse of the length scale at which the corresponding cosine is cut off, and the overall system ``realizes'' it. This is supported by the results of the rigorous RG equations describing the commensurate-incommensurate transition \cite{commIncommRG}. If we parameterize the short-distance cutoff as $\alpha=\alpha_0e^\ell$, where in each RG step $\ell$ increases incrementally, one may approximate the threshold at which the $g_{1,2}$ cosines are cut-off as
\begin{equation}
    \ell_{1,2}^{*}=\ln\frac{1}{\left(k_{F}\pm k\right)\alpha_{0}}\label{eq:ell12}.
\end{equation}
Generically, $\alpha_{0}\sim{\cal O}\left(k_{F}^{-1}\right)$, and one of the cosine terms, depending on the sign of $k$, will have a small $\ell^*$,  influencing the RG flow in only a negligible way. The important consequences of our theoretical model can be well understood even for $m_{\rm sec}=0$, and we leave the discussion on the role of the mass term in the RG flow to Appendix \ref{app:massRG}.

Finally, we derive the RG equations using operator product expansion (OPE) \cite{cardy_1996}. We define the dimensionless coupling constants $y_{s}=\frac{g_{s}}{\pi u_{s}}$, $y_{1,2}=\frac{g_{1,2}}{\pi \bar{u}}$, with $\bar{u}=\frac{u_c +u}{2}$. In the following we also neglect the flow due to velocity differences of the different sectors, which only impact the flow in higher orders. Our equations are thus
\begin{subequations}
\begin{equation}
    \frac{d}{d\ell}y_{s}=\left(2-2K_{s}\right)y_{s}-\frac{c_{1}\left(\ell\right)y_{1}^{2}+c_{2}\left(\ell\right)y_{2}^{2}}{4}\label{eq:RGys},
\end{equation}
\begin{equation}
\frac{d}{d\ell}K_{s}^{-1}=\frac{1}{2}y_{s}^{2}+\frac{c_{1}\left(\ell\right)y_{1}^{2}+c_{2}\left(\ell\right)y_{2}^{2}}{8},
\end{equation}
\begin{equation}
\frac{d}{d\ell}y_{1}=\left(2-\frac{\tilde{K_{c}}}{2}-\frac{K_{s}}{2}-\left(1-\frac{Q}{\sqrt{2}}\right)^{2}\tilde{K}-\frac{y_{s}}{2}\right)y_{1},\label{eq:RGy1}
\end{equation}
\begin{equation}
\frac{d}{d\ell}y_{2}=\left(2-\frac{\tilde{K_{c}}}{2}-\frac{K_{s}}{2}-\left(1+\frac{Q}{\sqrt{2}}\right)^{2}\tilde{K}-\frac{y_{s}}{2}\right)y_{2},\label{eq:RGy2}
\end{equation}
\begin{equation}
\frac{d}{d\ell}\tilde{K}^{-1}=\frac{c_{1}\left(\ell\right)y_{1}^{2}+c_{2}\left(\ell\right)y_{2}^{2}}{4},
\end{equation}
\begin{equation}
\frac{d}{d\ell}\tilde{K_{c}}^{-1}=\frac{c_{1}\left(\ell\right)y_{1}^{2}+c_{2}\left(\ell\right)y_{2}^{2}}{8}\label{eq:RGKc}.
\end{equation}
\end{subequations}
The scale dependent factors $c_1$ and $c_2$ impose a smooth cutoff of scale $\chi$ for the incommensurate cosines, where we use $c_{1,2}\left(\ell\right)=\frac{1}{2}\left[1-\tanh\left(\frac{\ell-\ell_{1,2}^{*}}{\chi}\right)\right]$. We note that we do not include in the RG flow generated bilinear cross terms such as $\partial_x \phi \partial_x \phi_c$, as feeding these back into the RG equations results in higher (third) order corrections for $y_{1,2}$. We stress once more that such modifications will not lead to a qualitative alteration of the phase diagram.

\begin{figure}
\includegraphics[scale=0.45]{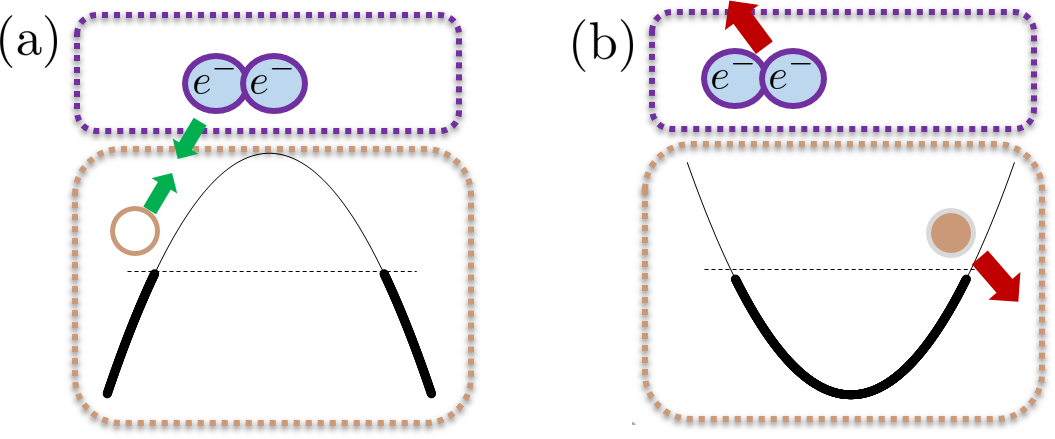}
\caption{\label{fig:Qkademonstration} The role of repulsive forward-scattering interactions between the primary electrons (highlighted by purple frame) and the secondary fermions (in the brown frame). (a) If the secondary sector is ``hole-like'', a secondary quasiparticle tends to bind two electrons to itself, thus the primary-secondary forward scattering interaction $Q$ enhances the $g_1$ term [Eq.~\eqref{eq:Hintboson}]. (b) For a ``particle-like'' sector, secondary fermions and electrons repel, leading to suppression of $g_2$ by $Q$. ($g_1$ and $g_2$ are coupling constants of backscattering interactions between the primary and secondary wires.) Conversely, if the interactions between the primary and secondary wires are attractive, the opposite of this scenario occurs, as can be inferred from the sign of $Q$ in Eqs.~\eqref{eq:RGy1}--\eqref{eq:RGy2}. }
\end{figure}

We proceed to make some observations regarding the derived RG equations. It is apparent from these equations that the forward-scattering primary-secondary interactions, embodied entirely by $Q$, ``favor''  $y_1$ over $y_2$, or vice versa. An intuitive understanding of this may be obtained by considering the secondary sector as comprised of either electrons with $k>0$, or holes with effectively  $k<0$ 
\footnote{Strictly speaking, the Fermi momenta values of electrons and holes with the same densities are identical, but the Fermi velocities are opposite.  This implies that electron states with positive velocities (right moving electrons) cross the Fermi energy at $+\left|k_F\right|$, while hole states with positive velocities (right moving holes) cross the Fermi energy at $-\left|k_F\right|$. We use this convention throughout the manuscript [see Eq.~\eqref{eq:bosonization}]}, 
see Fig.~\ref{fig:Qkademonstration}. The fugacity $y_1$ for example, describes the interaction between a secondary hole and an electron pair. It is thus clear why repulsive electron-electron interaction ($Q>0$) would favor it and decrease its scaling dimension (and why they would have the opposite effect  on $y_2$). Without loss of generality, we shall henceforth assume the secondary part was tuned such that $k$ is negative and in the vicinity of  $-k_F$, so we neglect $y_2$ and denote $y=y_1$, $\ell^*=\ell^*_1$.
The analysis for a setup with positive $k$ ("particle-like") can thus be extracted from the "hole-like" case by simply taking $Q\to-Q$. 

\subsubsection{Pairing and the competition between inter and intra wire repulsion}
The interplay between $y_s$ and $y_{1,2}$ reflects the competition between intra-wire repulsion and induced pairing due to inter-wire repulsion leading to polarization of the secondary wire. Examining the RG equations we observe that the fugacity $y_s$ starts out positive due the repulsive interaction, and is diminished by the secondary-mediated backscattering (the $y_1^2$ and $y_2^2$ terms in Eq,~\ref{eq:RGys}). If it crosses over to a negative value, the competition transforms into cooperation, as the two kinds of coupling grow (in their absolute value) together.

\begin{figure}
\includegraphics[scale=0.54]{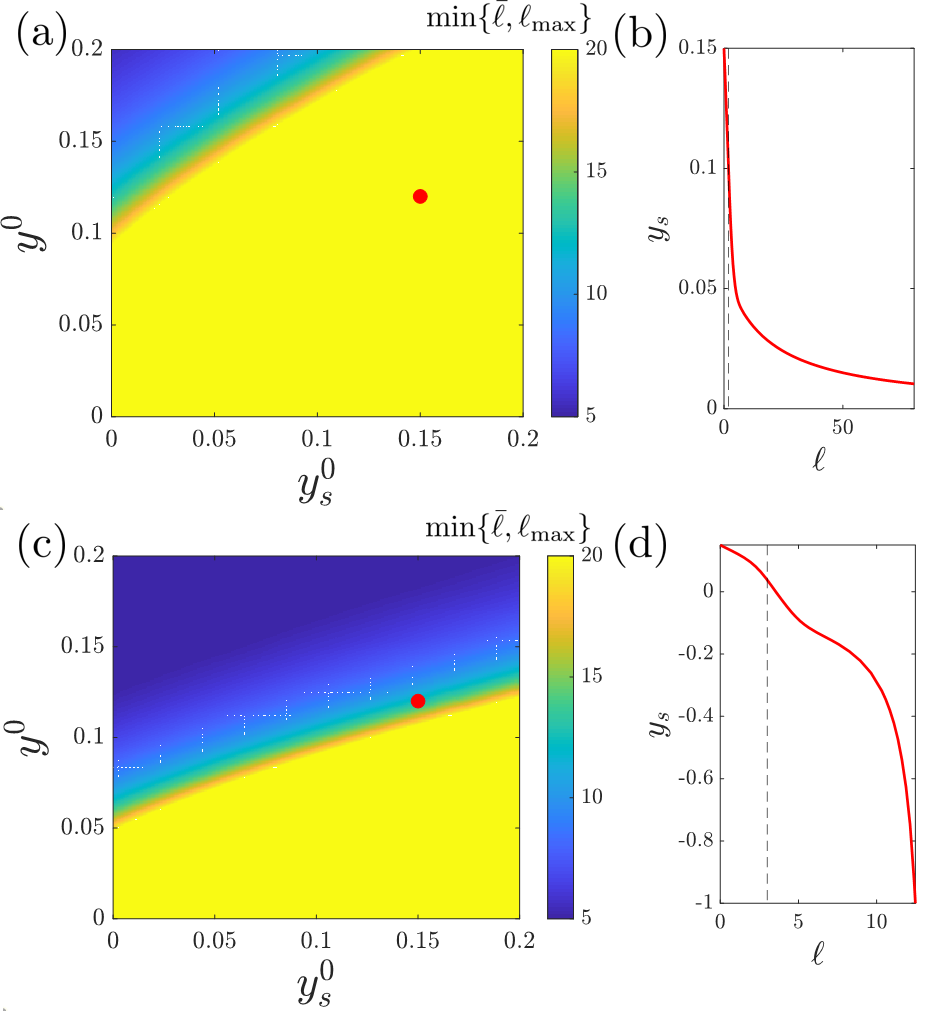}
\caption{\label{fig:2cutoofs} Pairing and the competition between inter and intra wire repulsion, and its dependence on the commensurability cutoff. The RG Eqs. \eqref{eq:RGys}--\eqref{eq:RGKc} are integrated up to a point where $y_s=-1$ (indicated by $\bar{\ell}$), or to an infrared cut off $\ell_{\rm max}$. (a) A plot of ${\rm min}\{\bar{\ell},\ell_{\rm max}\}$ as a function of the bare $y^0,y_s^0$.
$\bar{\ell}<\ell_{\rm max}$ signals the opening of a gap in the spin sector, which grows larger as $\bar{\ell}$ becomes smaller. In the opposite case, $y_s$ flows to weak coupling and the spin sector is gapless. $\bar{\ell}=\ell_{\rm max}$ then marks the phase boundary between the gapped and gapless spin phases.
Here we have set the cutoff due to the incommensurability to $\ell^*=2$. (b) Representative RG flow of $y_s$ (red) for bare values of $y_s^0=0.15$, $y^0=0.12$, indicated by the red dot in (a). The dashed line marks $\ell^*$. (c)--(d), the same as (a)--(b), with $\ell^*=3$. In all the plots we have $K_c^0=0.85$, $K^0=0.6$, $\upsilon=0.15$, $\chi=1$. We enforce $K_s,y_s$ to follow the SU(2) symmetric line, $K_s=1+\frac{y_s}{2}$.}
\end{figure}

An interesting situation arises if $y_{1,2}$ drives the system to a point where $y_s<0,$ and $K_s<1$, and is subsequently cut off by the incommensurability (or the mass term). Then, the terms $y_1^2$ and $y_2^2$ will no longer be in the the flow equations, and $y_s$ will flow to strong coupling according the formula:
$$y_{s}\left(\ell\right)\approx-\frac{\left|y_{s}\left(\ell^{*}\right)\right|}{1-\left(\ell-\ell^{*}\right)\left|y_{s}\left(\ell^{*}\right)\right|},$$ and a spin gap will open (with the charge sector remaining in a Luttinger liquid phase). The spin gap may be evaluated as $\Delta_s\approx\Lambda e^{-\bar{\ell}}$, with $\Lambda \sim v_s/\alpha_0$ a characteristic bandwidth and $y_s\left(\bar{\ell}\right)=-1$. This is illustrated by Fig. \ref{fig:2cutoofs}, where the $y,y_s$ competition and its dependence on $\ell^*$ is made clear. In particular, the phase boundary between the paired and gapless phases is highly affected by the incommensurate cutoff. We note that one may use Fig. \ref{fig:2cutoofs}a,c to extract an ``effective attraction'', by following iso-lines of $\Delta_s$ down to $y^0=0$ ending on a value $y_s^0<0$.

\subsubsection{Highly correlated secondary electrons}
The role of strong interactions in the secondary wire is made clear by Eqs. \eqref{eq:RGy1}--\eqref{eq:RGy2}. In the limit of vanishing interactions, $\tilde{K}_c,K_s,\tilde{K}\to1$, $Q\to0$, positive $y_s$ will always render the couplings $y_{1,2}$ irrelevant. Thus, a sufficiently small value of $K$, reflecting this strong repulsion of secondary fermions, is crucial to establish any of the non-trivial phases we mentioned. Notice that the same is true in principal for a small value of $\tilde{K}_c$. However, one should keep in mind that small bare $K_c$ is usually accompanied by large $K_s$ (and $y_s$), which counteracts the pairing mechanism we aim to activate. This can perhaps be circumvented by considering longer-range interactions in the primary electronic system, which affect mainly the value of $K_c$.

\subsubsection{Nearly perfectly tuned densities}
The final scenario we have yet to consider in the context of the RG flow is the one where $y$ reaches strong coupling \textit{before} $\ell$ exceeds $\ell^*$. This signals the approximate position of the commensurate-incommensurate transition (up to corrections due to the flow of $k_+$~ \cite{commIncommRG,Tsuchiizu2001}). As mentioned above, this is the phase where a secondary hole is bound by the interactions to a primary electron pair, whose formation originates in the opening of the spin gap.

\subsubsection{Pairing mechanism}\label{sec:pairingmechnism}
Let us briefly discuss the essential physics behind the pairing instability in the primary wire triggered by capacitive coupling to the secondary wire. The enhanced correlations between the two electron species in the primary system may be understood as a result of \textit{constructive interference} between backscattering events of secondary quasi-particles with spin-$\uparrow$ and spin-$\downarrow$ primary electrons, which are in-phase with one another.

To elucidate this point, consider two kinds of backscattering interactions contained within ${\cal H}_{\rm int}$ [Eq.~\eqref{eq:Hint}]
\begin{equation}
    g_{\uparrow}c_R^\dagger c_L \psi_{L\uparrow}^\dagger \psi_{R\uparrow}+{\rm h.c.,}\,\,\,\,\,g_{\downarrow}c_R^\dagger c_L \psi_{L\downarrow}^\dagger \psi_{R\downarrow}+{\rm h.c.}
\end{equation}
In the symmetric setup we analyze, $g_\uparrow=g_\downarrow$. To illustrate the importance of constructive interference between the terms, consider a different scenario where the two coupling constants have a relative phase of $\theta_{\uparrow\downarrow}$ and equal magnitude. In such a scenario, the RG equations may be modified in quite a drastic way.  Most importantly, if we examine the flow of $y_s$ [Eq.~\eqref{eq:RGys}] we find the the term proportional to $y_{1/2}^2$ acquires a prefactor of $\cos\theta_{\uparrow\downarrow}$. We reiterate that this term is responsible for reducing the primary intra-wire repulsion, eventually leading to \textit{attraction} later in the RG flow. Thus, one finds that a relative phase between the backscattering sectors may hinder the pairing.  

As a particularly useful extreme example, consider $\theta_{\uparrow\downarrow}=\pi$, i.e., $\uparrow$-electrons are repelled by the secondary quasi-particles, while $\downarrow$-electrons are attracted to them (or vice-versa). The cosine will then reverse the trend we find in Eq.~\eqref{eq:RGys}, and make the intra-wire repulsion more significant at lower energy scales.

Examining the effect of such a fictitious phase illustrates how the interactions of primary electrons with an additional mutual element (the secondary excitations) promote inter-species correlations and eventually pairing, regardless of the sign of the inter-wire interactions. Simply put, electrons tend to backscatter simultaneously in both species, even if they repel one another. The competition between their tendency to repel and their backscattering-induced correlations will determine the fate of the system.

It is worth emphasizing that the basic mechanism we describe here, the in-phase backscattering, does not necessarily require the secondary quasi-particles to be ``hole-like''. Though having secondary holes (instead of electrons) will make the primary wire more susceptible to the pairing instability [through physically attracting the primary electrons and modifying the scaling dimensions in Eqs.~\eqref{eq:RGy1}--\eqref{eq:RGy2}], the tendency towards inter-species correlations does not rely on it.

\subsection{\label{sec:phasediagram} Phase diagram}
\begin{figure*}
\includegraphics[viewport=67.1786bp 0bp 1131bp 324bp,scale=0.55]{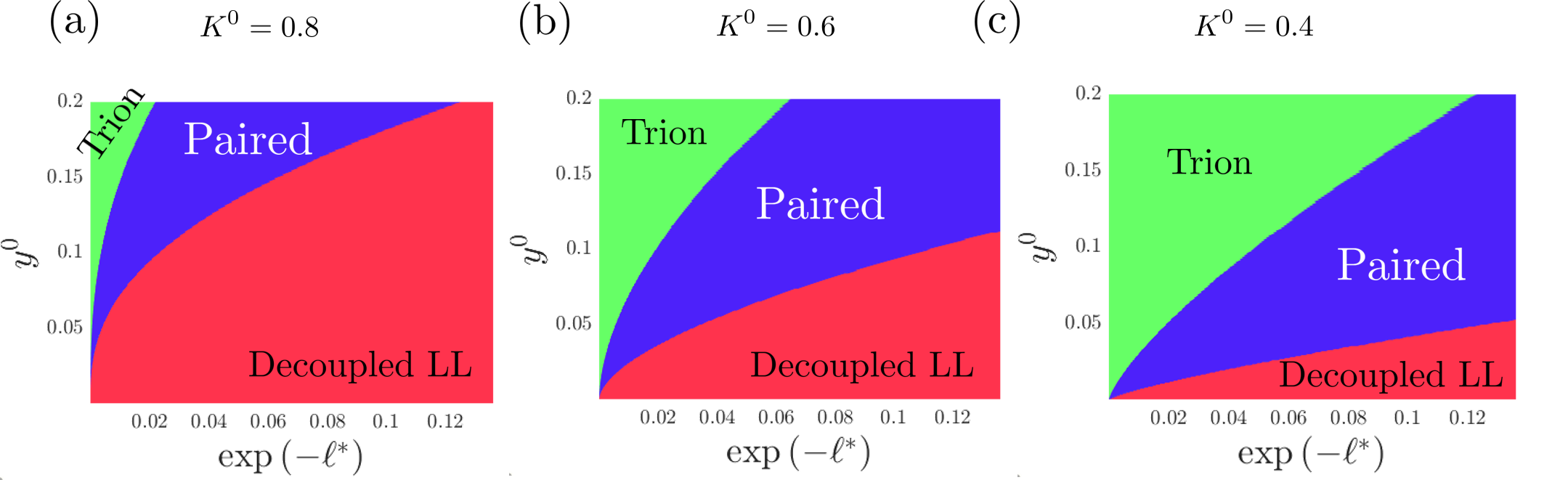}
\caption{\label{fig:phasesdiffKa} Zero temperature phase diagram calculated from integrating the RG equations, for different values of secondary fermions interaction strength, $K^0$, [indicated above each panel in (a)--(c)]. Notice that the $x$ axis, ${\rm exp\left(-\ell^{*}\right)}\propto\left(k_{F}+k\right)\alpha_{0}$, Eq.~\eqref{eq:ell12}, for $k_F=-k$ the densities in the wires are commensurate . The decoupled LL phase is defined by the flow of the spin fugacity of the primary wire, $y_{s}$, to zero. The trion phase is where $y$, the inter-wire back-scattering fugacity, reaches strong coupling for $\ell<\ell^{*}$. In the spin-gap paired phase, $y_{s}$ reaches strong coupling (and is negative). In this figure we set the bare spin fugacity $y_{s}^{0}=0.1$, the bare  Luttinger parameter of the primary charge sector  $K_{c}^{0}=0.9$, the forward scattering interaction between the wires $\upsilon=0.12$ [see before Eq.~\eqref{eq:KcTransformed}], and the scale determining the smoothness of the incommensurability cutoff in the RG process $\chi=1.5$ [see following Eq.~\eqref{eq:RGKc}].}
\end{figure*}

We are now in the position to examine which of the three possible phases is established as a function of experimentally controllable parameters, namely the incommensurability cutoff $\ell^*$ (controlled by, e.g., a gate voltage applied to the secondary wire), and the interwire bare interaction with high momentum transfer $y^0$ (typically tuned by the distance between the primary and secondary parts). This is explored in Fig.~\ref{fig:phasesdiffKa}, where the different phase boundaries are found. We observe that moving towards the commensurate point (increasing $\ell^*$) tends to favor the ``locked'' trion phase over the paired one.
The role of the Luttinger parameter $K$ is also apparent: as it gets smaller (reflecting stronger interaction in the secondary system), the region in phase space where one can easily tune between the non-trivial phases is greatly enhanced.

\begin{figure}
\includegraphics[scale=0.61]{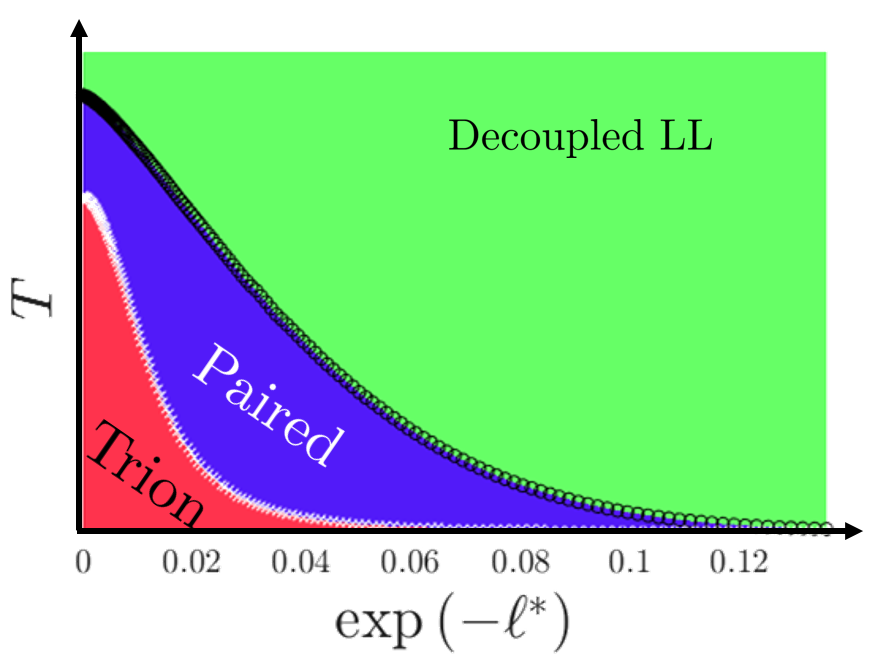}
\caption{\label{fig:Tdiagram} Schematic finite temperature phase diagram, as a function of $\ell^{*}$. The phase boundary of the decoupled-paired phases is determined by $e^{-\bar{\ell}}$. The boundary for the trion phase is set by $e^{-\bar{\bar{\ell}}}$, multiplied by $c\left(\bar{\bar{\ell}}\right)$, as to account for the incommensurate cutoff. Parameters used for generating this plot: $y_{s}^{0}=y_{1}^{0}=0.1$, $K_{c}^{0}=0.85$, $K^{0}=0.4$, $\upsilon=0.1$, and $\chi=1.5$.}
\end{figure}

The presented RG analysis allows us to deduce a schematic finite temperature phase diagram, see Fig.~\ref{fig:Tdiagram}. As a function of a single parameter, e.g., the incommensurability cutoff $\ell^{*}$, we calculate the RG scale at which $y_{s}$ reaches strong coupling $\bar{\ell}$, as well as the scale corresponding to the pinning of the $y$ term, $\bar{\bar{\ell}}$. We estimate the gaps associated to these coupling constants as $\propto  \Lambda e^{-\bar{\ell}}, \Lambda e^{-\bar{\bar{\ell}}}$, allowing us to determine an appropriate energy scale for the transition. It can be seen that for a certain range of the detuning parameter (depending on interaction strength, etc.), as we lower the temperature the system may experience a crossover from the Luttinger liquid phase, through the paired Luther-Emery liquid, and finally to a trion phase where the primary pairs are bound to secondary fermions.

We now turn to discuss the properties of the system in its three possible phases.

\subsubsection{\label{decuopledphase} Decoupled phase}

In the decoupled phase, where no spin-gap is opened for the primary electrons, the electronic system has the usual features of a spinful Luttinger liquid, e.g., spin-charge separation and a logarithmic tendency towards a $2k_F$ spin-density-wave (SDW) formation \cite{giamarchi2004quantum}. As mentioned before, its properties are slightly modified by the bilinear interaction terms with the secondary fermions. 
We focus our attention below on the two primary-gapped phases.

\subsubsection{\label{spingapphase} Paired phase}
When a finite spin-gap occurs, but when the secondary sector $\phi$ is still gapless, we find generically a tendency towards a charge-density-wave (CDW) phase in the primary wire, with sub-dominant superconducting (SC) fluctuations. To see this quantitatively, consider the susceptibilities of these two kinds of order parameters,
\[\chi_{\rm CDW}\sim\omega^{\left( \tilde{K}_c + Q^2\tilde{K} + K_s -2\right)},\]
\[\chi_{\rm SC}\sim\omega^{\left( \tilde{K}_c^{-1}  + K_s -2\right)},\]
with $\omega$ the energy scale. When the spin is gapped, we should take $K_s\to0$. If we further assume that $Q\approx0$ it is clear that repulsive interactions will favor CDW susceptibility, as it has a negative exponent with a value that is larger than the exponent of the SC susceptibility. Interestingly, strong primary-secondary interaction (which increases the parameter $Q$) can render the SC pairing fluctuation more dominant if
\begin{equation}
    \tilde{K}_c + Q^2\tilde{K}>\tilde{K}_c^{-1}.\label{eq:scCondition}
\end{equation}
Thus, for sufficiently weak interactions between the spinful electrons ($K_c \rightarrow 1$), $Q$ drives the spin-gapped system to the 1D phase analogous to a superconductor, i.e., dominant quasi-long-range order for pair-pair correlations (this also happens for bare attractive interactions, but here it was obtained from repulsion only). The discussion above ultimately reveals that inducing a spin gap that pairs electrons in the primary 1D system does not necessarily mean one has induced \textit{superconductivity}. The details of the interactions will thus determine the overall phase.

Several measurements can be experimentally performed in order to study the properties of the system and distinguish the different phases.
While in the decoupled phase we expect to observe power-law behavior in single-electron tunneling experiments to the primary wire, the spin-gapped paired phase should display a gap in the tunneling spectra~\cite{Voit_1996}.

As the charge mode remains gapless, two-terminal conductance measurement of such a clean system should still yield $G=2\frac{e^{2}}{h}$. We note that for the case of a single impurity in the system, one may detect a conductance \textit{drop} when entering the paired phase. This is due to the effective $K_{s}\rightarrow0$, compared to $K_{s}\rightarrow1$ in the gapless case. For an impurity of strength $g_{{\rm imp}}$, the energy scales determining the scaling of the conductance for the decoupled LL and paired phases are \cite{KaneFisher,KaneImpurity}
\[
E_{{\rm LL}}\approx \Lambda\left(\frac{g_{{\rm imp}}}{\Lambda}\right)^{\frac{2}{1-K_{c}}},\,\,\,E_{{\rm sG}}\approx \Lambda\left(\frac{g_{{\rm imp}}}{\Lambda}\right)^{\frac{2}{2-K_{c}}},
\]
respectively. Since $g_{{\rm imp}}<\Lambda$, we find $E_{{\rm sG}}>E_{{\rm LL}}$ and a lower conductance for the paired phase is expected. We further expect that in the presence of an impurity the difference in conductance between the paired and decoupled phases should be most pronounced at intermediate energy scales, i.e., around $E_{{\rm LL}} \lesssim \omega \lesssim E_{{\rm sG}}$.

\subsubsection{\label{lockedphase}  Trion phase}
Since this phase is not the main focus of our work, we discuss here only its important features, and relegate additional technical details to Appendix~\ref{app:lockeddetails}.  We assume that the spin has already been gapped out, such that $\left\langle \cos\left(\sqrt{2}\phi_{s}\right)\right\rangle$ is finite, and change the basis to
\begin{equation}
    \phi_{g}=\frac{\phi_{c}+\sqrt{2}\phi}{\sqrt{3}},\,\,\,\,\,\phi_{f}=\frac{\sqrt{2}\phi_{c}-\phi}{\sqrt{3}},\label{eq:lockedsectors}
\end{equation}
and the same transformation for the dual variables $\theta_{g,f}$ is implicit. The bosonic field  $\phi_g$ is pinned by the interaction term $\propto g_1 \cos\left(\sqrt{6}\phi_g\right)$ [Eq.~\eqref{eq:Hintboson}], and the remaining sector combining charge modes in the primary and secondary wires is described by a LL Hamiltonian
\begin{equation}
    {\cal H}_{f}=\frac{u_{f}}{2\pi}\left[\frac{1}{K_{f}}\left(\partial_{x}\phi_{f}\right)^{2}+K_{f}\left(\partial_{x}\theta_{f}\right)^{2}\right],
\end{equation}
with \[K_f=\sqrt{\frac{2u_{c}K_{c}+u K-\sqrt{8}\frac{V_\theta}{\pi}}{\frac{2u_{c}}{K_{c}}+\frac{u}{K}-\sqrt{8}\frac{V_\phi}{\pi}}},\] where $u_f$ is given in Eq.~\eqref{eq:appBparameters}, and we relaxed the assumption of Eq.~\eqref{eq:specialV} for this expression.

To gain physical insight as to this gapless sector, we consider the operator $\Psi_{\rm trion}\equiv\sum_{r,r',r''}\psi_{r\uparrow}\psi_{r'\downarrow}c^\dagger_{r''}$, which creates a trion of spin-singlet electron pair in the primary wire and a hole in the secondary wire. Bosonizing this operator, one finds that when $\phi_s,\phi_g$ are pinned,
\begin{equation}
    \Psi_{{\rm trion}}\left(x\right)\sim e^{i\sqrt{3}\theta_{f}}\cos\left(\frac{1}{\sqrt{3}}\phi_{f}-k_{F}x\right).\label{eq:trionbosonized}
\end{equation}
By comparing Eq.~\eqref{eq:trionbosonized} to the operators of a sector of 1D spinless fermions, we infer that the point $K_{f}=3$ corresponds to that of free non-interacting fermions (trions). This implies that for $K_{f}<3$ the system will be dominated by CDW correlations of these trions, $O_{{\rm CDW}}^{{\rm trion}}=\Psi_{{\rm trion}}^{\dagger}\left(x\right)\Psi_{{\rm trion}}\left(x\right)$, whereas pairing correlations given by the operator $O_{{\rm SC}}^{{\rm trion}}=\Psi_{{\rm trion}}\left(x\right)\Psi_{{\rm trion}}\left(x+a\right)$ will dominate if $K_{f}>3$. Additionally, operators corresponding to local impurity backscattering which impact the secondary or primary systems will be irrelevant for $K_{f}>3$ in this phase.

We should point out that $K_{f}\geq3$ corresponds to a rather large attraction in the $\phi_{f}$-sector, i.e., between nearby trions. Physically, this would mean that $V_\phi$, responsible for the attraction of secondary holes and primary electrons, overwhelms the intra-wire electron-electron and hole-hole repulsive interactions. Thus, although this ``superconducting-trions'' phase exists in the phase diagram of our model, it corresponds to a somewhat extreme unphysical limit of realistic setups.

Similarly to the paired phase we discussed earlier, a single-particle gap exists in the spectral function of the primary electronic wire, but also in that of the secondary system. Any single-particle tunneling operator to either of the two 1D systems contains some exponent of the $\theta_g$ and/or $\theta_s$, which are completely disordered, and thus render the tunneling irrelevant.

\begin{table}
\begin{centering}
\bgroup\def\arraystretch{2}
\begin{tabular}{|c||c|c|c|}
\cline{2-4} \cline{3-4} \cline{4-4}
\multicolumn{1}{c|}{} & \textbf{Decoupled} & \textbf{Paired} & \textbf{Trion}\tabularnewline
\hline
Single-electron tunneling & \textit{Power-law} & \textit{Gap} & \textit{Gap}\tabularnewline
\hline
Two-terminal conductance $\left[\frac{e^2}{h}\right]$ & $2$ & $2$ & $4/3$\tabularnewline
\hline
\end{tabular}
\egroup
\caption{Discerning the decoupled LL phase, spin-gap paired phase, and the trion phase. Single electron tunneling into the primary wire is expected to vanish with a power-law at zero bias or to have a finite gap. The two-terminal conductance of the primary wire is measured with the secondary system grounded.}
\par
\end{centering}
\end{table}

In order to distinguish these phases one may measure the two-terminal electrical conductance of the primary wire while keeping the secondary one grounded. Since the remaining sector carries a fraction of the total charge, one should expect to find fractional conductance. From a straight-forward calculation \cite{fractionalGSYO}, one finds the ``ideal'' conductance of a clean system measured in this setup is $G=\frac{4}{3}{e^2}/{h}$. This contrasts the predicted $G=2{e^2}/{h}$ in the two other phases where $\phi_c$ remains gapless. 
We emphasize that this fractional conductance is not due to current measurement in only one part of the system. In such a setup there is a counterflow of current in the secondary wire, yet it is also fractional, corresponding to a cross-conductance of  $-\frac{2}{3}{e^2}/{h}$. Thus, the total current in the system remains fractional (see Appendix~\ref{app:fractionalconductancecalc}).
We note that unlike the scenarios considered in Ref. \cite{fractionalGSYO}, the fractional conductance here is expected to be a much more robust feature. This is because $g_{1,2}$ are not generated by higher-order backscattering processes, but rather they are first-order in the interaction strength.

The measurement of the fractional conductance in this phase of the system can provide an important experimental tool for tuning the system into the (possibly more interesting) paired phase. By monitoring the conductance of the primary channel, one can modify the chemical potential of the secondary $c$-fermions to a point where the fractional conductance emerges. Then, one can detune the chemical potential back, right up to where the conductance returns to its integer value. Per our phase diagram, the total system should generically end up in the paired spin-gap region, unless the interwire interactions are too weak.

\section{\label{sec:previousWorks} Extensions}

\subsection{\label{sec:Littlemodel} Connection to Little's model}
Inducing attraction between electrons mediated by extrinsic repulsive interactions of these electrons and an auxiliary system was originally proposed by Little \cite{LittleWA}. Little considered a one-dimensional system coupled by Coulomb interactions with strength $V$ to local polarizable side-chains along the system (that was termed ``spine''). Excitations of these so-called polarizers, which cost a finite energy $\tilde{E}$, mediate attractive interactions between system electrons, proportional to $\sim\frac{V^{2}}{\tilde{E}}$. This attraction then competes with the intrinsic electron-electron repulsion in the chain hosting the electrons.

The model Little put forward was studied more carefully by Hirsch and Scalapino (HS) \cite{HirschScalpainoPRL,HirschScalapino}. HS studied this model by employing perturbation theory in the weak and strong electron-polarizer coupling limits, as well as using numerical methods, namely small-scale quantum Monte-Carlo. HS found certain regimes of parameters where superconductivity occurs in such a system, yet concluded that $2k_{F}$ CDW of pairs generally has larger susceptibility, even when pairing occurs. This is reminiscent of the conclusions we draw in Sec.~\ref{sec:phasediagram} for the setup proposed here, which similarly has a SC order or CDW of pairs in the spin gapped paired phase.

At first glance, our model might seem to have very little to do with Little's model. The pairing mechanism we discuss in this work relies on interaction with secondary fermions, which are either in a Luttinger liquid phase or gapped (such that the gap does not overwhelm the interactions with the electronic system). In fact, Little's model presented in Ref.~\cite{LittleWA} can be shown to coincide with a certain limit of the model we presented here. It corresponds to extremely localized fermions, i.e., $v_{\rm sec}\rightarrow0$ in our description, which are gapped, with the energy required to polarize/excite one of the side-chains polarizers being the gap. A solitonic excitation in our version of secondary fermions Hamiltonian, Eq. \eqref{eq:Habososn}, is the analog of an excited polarizer (in a very non-itinerant limit).

\begin{figure}
\includegraphics[scale = 0.3]{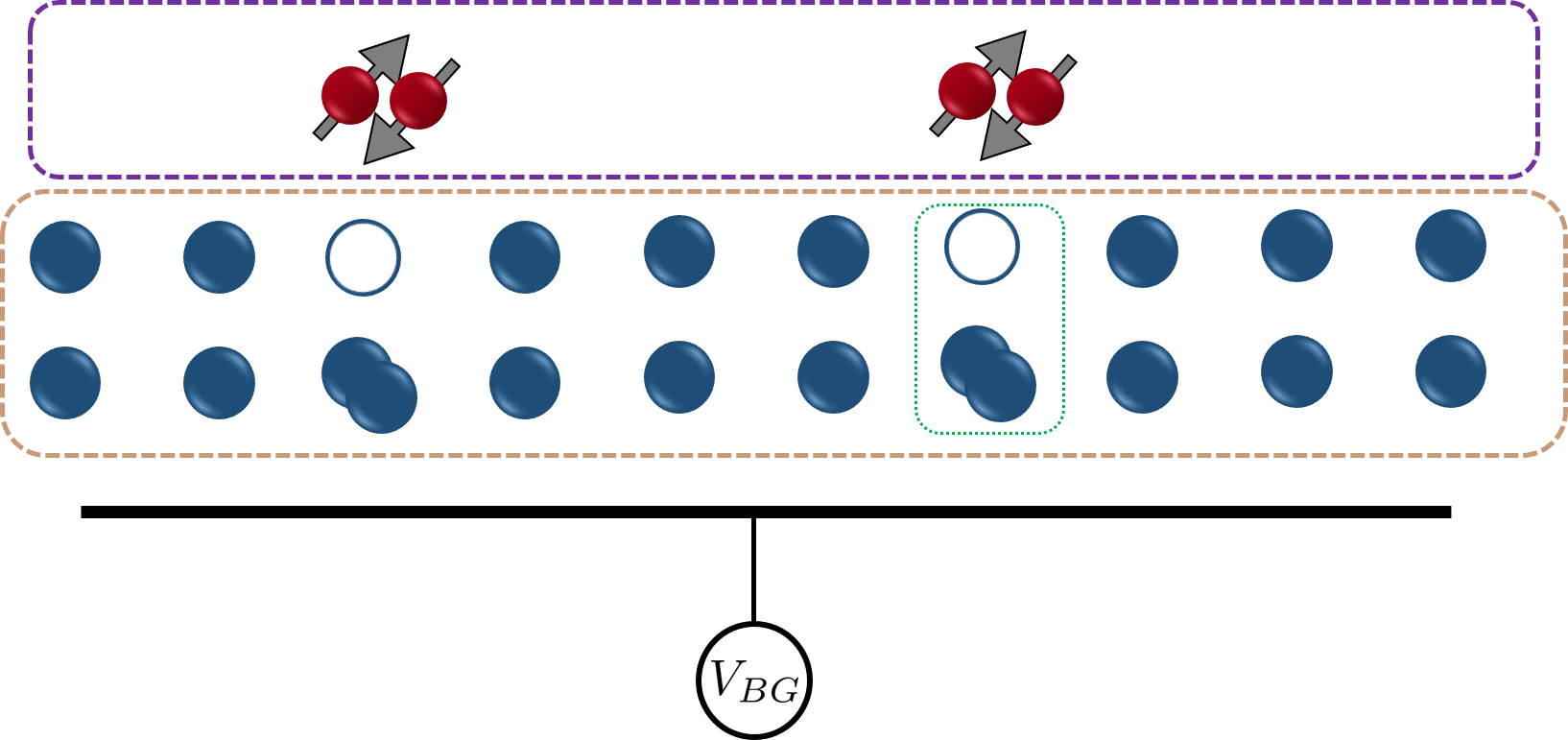}
\caption{\label{fig:littlemottschematics}
Schematic proposal of a Mott insulator as the secondary pairing mediator. A primary wire (dashed purple frame) is capacitively coupled to a two-leg Hubbard ladder (dashed brown frame) at half filling (full circles mark electrons and empty ones are vacancies). The pairing is mediated by coupling to the $\phi_{\rho-}$  sector (presenting the charge difference between the two legs) of the Mott insulator (see text).
A potential difference between the two legs may be induced by a back-gate voltage $V_{BG}$ as an experimental knob which controls the proliferation of solitons [the doublon-holon excitation highlighted in a dotted green line in the figure]  in this sector. This setup bears some resemblance to the model by Little \cite{PairingLittle1981}, where the two polarizer states are replaced by the rung ground and excited state (the latter being a doublon-holon excitation).
}
\end{figure}

To better illustrate the connection between Little's proposal and some particular sector of our model, we consider one possible implementation of a secondary system that may be employed: a two-leg Hubbard ladder exactly at half-filling, illustrated in Fig. \ref{fig:littlemottschematics}. For our purposes, it embodies a minimal representation of a strongly correlated Mott insulator. Strong interactions pin the system into a state of one electron per site, with a charge gap proportional to the amplitude of the on-site interactions. We assume that the Hubbard ladder is in its arguably most generic phase, the so-called ``D-Mott'' phase, or the rung singlet, appropriate when the on-site interactions are large compared to the nearest-neighbor repulsion and the exchange energy~\cite{DMOTT}. In this phase, all charge and spin sectors of the system are gapped.

We now consider the ``relative-charge'' sector of the ladder, i.e., the one that measures the total difference in charge between the different legs, $\phi_{\rho-}\propto\phi_{\rho,1}-\phi_{\rho,2}$, where $\phi_{\rho,i}$ represents the charge sector of the $i$'th leg, and argue that it is the most important sector in terms of its effects on the nearby electronic system we wish to induce pairing in. The interaction-induced gap in this sector is considerably smaller compared to that of the total charge sector, since it has a smaller effective interaction parameter ($K_{\rho-}$ is much closer to 1, whereas $K_{\rho+}$ can be rather small, as it accounts for the total charge sector).

Assuming the dominant interaction of the primary electronic system is with the leg closest to the system (the leg with $i=1$), the interactions between the primary spinful wire (described by the sectors $\phi_c$ and $\phi_s$) and the Hubbard ladder secondary system are accounted for by
\begin{align}
    &\frac{V}{\pi^{2}}\partial_{x}\phi_{c}\partial_{x}\phi_{\rho,1}\nonumber\\
     & +\frac{2\tilde{g}_{1}}{\pi^{2}a^{2}}\cos\left(\sqrt{2}\phi_{s}\right)\cos\left(\sqrt{2}\phi_{c}+\sqrt{2}\phi_{\rho,1}-2k_F-2k_1\right)\nonumber\\
     & +\frac{2\tilde{g}_{2}}{\pi^{2}a^{2}}\cos\left(\sqrt{2}\phi_{s}\right)\cos\left(\sqrt{2}\phi_{c}-\sqrt{2}\phi_{\rho,1}-2k_F+2k_1\right),\label{eq:LittleHint}
\end{align}
with $k_F$ and $k_1$ corresponding to the Fermi momenta of the primary wire and closest Hubbard leg, respectively.
We have also absorbed a factor of $\left\langle \cos\left(\sqrt{2}\phi_{\sigma,1}\right)\right\rangle$ into the definition of the interaction coupling constants $\tilde{g}_{1,2}$,
with $\phi_{\sigma,1}$ corresponding to the spin sector of the leg interacting with the electron wire.
Assuming that the total charge sector is gapped, such that $\phi_{\rho+}$ is strongly pinned to its semi-classical strong-coupling value, we arrive at an expression similar to Eq. \eqref{eq:Hintboson}.
One important difference is that the $\phi_{\rho-}$ appears inside the cosine with a smaller prefactor, for example in $~\tilde{g}_1 \cos\left( \sqrt{2}\phi_c+\phi_{\rho,1} \right)$ (1 instead of $\sqrt{2}$).
This in turn makes the non-linear perturbations in Eq.~\eqref{eq:LittleHint} significantly more relevant as compared to those in Eq.~\eqref{eq:Hintboson}. (This is not surprising, as we have assumed that some of the degrees of freedom in the secondary ladder are already frozen by the interactions in the Mott insulator.)

A chemical potential difference between the two legs of the ladder can be generated, for example with back gate potential, leading to a term $\sim\mu^{*}\partial_{x}\phi_{\rho-}$. This will allow one to experimentally control $\ell^{*}$ (as it controls the commensurability condition) as well as the mass gap in such a setup, and tune it towards the desired phase for the overall system.

Let us briefly comment on the nature of the $\phi_{\rho-}$ solitons. As mentioned, in the D-Mott phase the density is one electron per site, with the spins forming a rung-singlet. A relative-charge particle (hole) would correspond to an electron being displaced from leg 2 to leg 1 (or vice-versa), without perturbing the spin-singlet nature of the electrons on the rung. Notice that such an excitation costs a finite amount of energy, approximately proportional to the difference between the on-site repulsion and inter-leg repulsion energies in the strong coupling limit. The analogy to Little's model now becomes a bit more transparent: the ground state and excited state of the polarizers correspond to a rung with one electron per site and a $\phi_{\rho-}$ hole, respectively. A key difference in our proposed setup is that we can account for interactions between these secondary fermions, and that we allow for their excitations to delocalize. These two effects, as can be inferred from our RG analysis and results, can be quite important in determining and manipulating the phase diagram.

On top of  allowing us to make a concrete connection to  Little's model, this specific realization of the secondary wire has additional significance. The discussion above suggests that under certain conditions, proximity of a spinful wire to a Mott insulator may result in pairing of electrons in that 1D system. Realistically, this Mott insulator does not necessarily have to be one-dimensional:  If its gaps are sufficiently larger than the interaction with the primary wire, the physics may be well-described by that of the insulator sites nearest to (and strongest interacting with) the electronic wire. Thus, our toy-model may be an appropriate effective description of a simple experimentally accessible setup, namely placing a nanowire on top of a strongly correlated Mott insulator.

\subsection{\label{sec:twoD} Higher dimensions}
The model we have presented here concerns the pairing one may induce in a spinful one-dimensional system, even when it has non-negligible intrinsic repulsive interactions. Due to the Mermin-Wagner theorem, ``true'' superconductivity cannot occur in a 1D system. Moreover, we have discussed superconducting tendencies brought on by our setup, and shown that they were generically expected to be sub-dominant to CDW order developing in the wire (whether for spin-singlet pairs or the so-called trions). We would therefore like to consider the extension of the setup presented here to a higher dimension.

\begin{figure}
\includegraphics[scale=0.57]{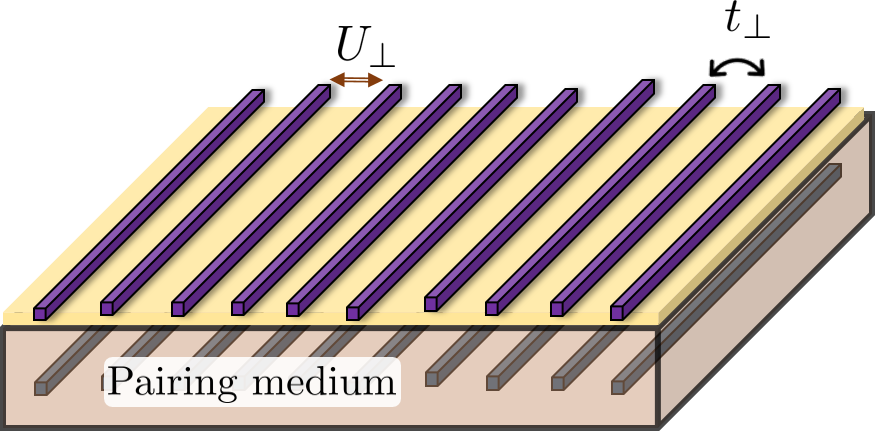}
\caption{\label{fig:wireconstr} A 2D extension of the model presented in this work. A quasi-1D lattice composed of an array of primary spinful wires is proximitized to a secondary ``pairing medium''. The latter may consist of separate tunable 1D systems hosting spinless fermions, or a strongly correlated Mott insulator, see the discussion in Sec.~\ref{sec:Littlemodel}. The wires are separated from the pairing medium by either an insulating layer (yellow) or by sufficient distance, as to not allow hopping of particles between the two primary and secondary parts. The inter-wire repulsive interaction $U_\perp$ and hopping amplitude $t_\perp$ are indicated.}
\end{figure}

Let us consider an array of spinful quantum wires, as depicted in Fig. \ref{fig:wireconstr}, each with its own secondary system, all decoupled from one another. This is a very anisotropic limit of a 2D system, ubiquitous in many coupled wires studies (c.f. Ref.~\cite{KaneLubensky2002}). One can then tune each such individual primary-secondary duo to the paired phase by changing the relevant controllable parameters, and allow weak hopping between neighboring primary wires, such that the tunneling amplitude $t_{\perp}$ is smaller than the induced spin gap $\Delta_{s}$. [We neglect the tunneling between adjacent secondary wires, which is justified if: (i) Strong interactions within them makes such tunneling irrelevant, (ii) each secondary system is individually tuned such that there is a Fermi momentum mismatch between neighboring secondary wires, or (iii) the secondary wire has a single particle gap rendering such hopping non-important (such a case is discussed in Sec. \ref{sec:Littlemodel}).]

For the sake of completeness, the remainder of this Section mostly recapitulates results discussed in Refs.~\cite{Smectic,DimensionalCrossoverHighTc,SpinGapProximityEffect}, in the terminology used in Sec.~\ref{sec:mainsec}. Namely, we assume the spin-gapped paired phase in the primary wires (induced solely by repulsion) and discuss the conditions under which it leads to long-range phase coherent superconductivity.

Since $t_\perp  \ll \Delta_s$ we can restrict our discussion to higher order processes of  ${\cal O}(t_{\perp}^{2}/\Delta_s)$ between adjacent (primary) electron wires.
The key points in this discussion will be best illustrated by considering just two spinful spin-gapped wires, with labels 1,2. We note that the results we outline below are somewhat different than those discussed in Ref.~\cite{FinkelsteinLarkinTwoTLL} for a similar setup, as large-momentum backscattering events (of the kind that lead to the formation of the pairing spin gap) were not considered in~\cite{FinkelsteinLarkinTwoTLL}. Defining the sectors $\phi_{\alpha\pm}=\frac{\phi_{\alpha1}\pm\phi_{\alpha2}}{\sqrt{2}}$, with $\alpha=c,s,$ the intra-wire charge and spin sectors, we write the Hamiltonian density of the two-particle hopping processes as
\begin{equation}
    {\cal H}_{J}\sim t_{\perp}^{2}/\Delta_s\cos\left(2\theta_{c-}\right)\cos\left(\sqrt{2}\phi_{1s}\right)\cos\left(\sqrt{2}\phi_{2s}\right),
\end{equation}
\begin{equation}
    {\cal H}_{p-h}\sim t_{\perp}^{2}/\Delta_s\cos\left(2\theta_{s-}\right)\cos\left(2\phi_{c-}\right),
\end{equation}
which are the Josephson and particle-hole couplings, respectively. The intra-wire spin-gaps render ${\cal H}_{p-h}$ irrelevant, as it contains a cosine of variables dual to the gapped ones, whereas ${\cal H}_{J}$ is relevant in the RG sense if the Luttinger parameter corresponding to the relative charge sector is not too small, i.e., $K_{c-}>\frac{1}{2}$. (This is because in the spin-gapped regime we replace cosines of the individual $\phi_{1/2s}$ by their expectation values.) Interestingly, this condition does not exclude very strongly interacting systems, as long as electron-electron interactions between wires is not too small compared with the intra-wire one. In other words, inter-wire repulsion actually makes ${\cal H}_{J}$ more relevant~\cite{SpinGapProximityEffect}. Presumably, ${\cal H}_{J}$ can now flow to strong coupling, with the $\theta_{c-}$ term inducing inter-wire phase coherence throughout the system, causing a Kosterlitz-Thouless transition into the superconducting regime, similarly to the studied phenomena in Refs. \cite{Smectic,DimensionalCrossoverHighTc}.

In fact, it is well established that in considering such setups one should include an additional term, proportional to the inter-wire interaction $U_{\perp}$,
\begin{equation}
    {\cal H}_{\pi{\rm CDW}}\sim U_{\perp}\cos\left(2\phi_{c-}\right)\cos\left(\sqrt{2}\phi_{1s}\right)\cos\left(\sqrt{2}\phi_{2s}\right).
\end{equation}
Comparing it to ${\cal H}_J$, a clear competition between pinning of the dual variables $\phi_{c-}$ and $\theta_{c-}$ is evident.
The $\pi$-CDW interaction triumphs over the Josephson pair hopping, i.e., more relevant in an RG sense, if $K_{c-}<1$, facilitating a $\pi$-CDW order in the higher dimensional system \cite{Smectic}.
This term is typically neglected when considering, e.g., stripes in high-$T_{c}$ superconductors, since it is assumed that each stripe can effectively have different carrier densities compared to its neighbors. This in turn causes spatial oscillation in the relative charge cosine term,
\[\cos\left(\sqrt{2}\phi_{c-}\right)\to\cos\left(\sqrt{2}\phi_{c-}+\delta x\right),
\]
with $\delta$ being proportional to the density difference between neighboring primary wires. On length scales much larger than $\delta^{-1}$, the $\pi$-CDW is thus rendered irrelevant.

In the absence of this kind of dephasing, superconductivity would prevail only if $U_{\perp}$ is sufficiently small as compared to $\frac{t_{\perp}^{2}}{\Delta_{s}}$, or if $K_{c-}$ is large enough. The latter requires that the long-wavelength part of the inter-wire repulsive interaction be rather large. Conversely, engineering a fully tunable quasi-1D system, each having a separate spin-gap, the CDW tendency may be greatly diminished (dephased) by modulating the electronic densities between adjacent wires, effectively sending ${\cal H}_{\pi{\rm CDW}}\rightarrow0$ at low enough energy scales.

We note here that the discussion above, which centered around coupling of two spinfull wires in the paired (spin-gap) phase, alludes to the fact that introducing further complexity to the primary system may lead to dominant superconducting tendencies even in 1D setups. As an example, one may consider CNTs, which host two spinfull sectors (one in each ``valley'')~\cite{RevModPhyscnt,dresselhaus1998physical,dresselhaus1998physical}. Manipulating the properties of such CNTs, one may induce pairing and superconducting tendencies by proximitizing them to a proper secondary system.

To conclude this section, the role of the scheme proposed in this work and explored in Sec. \ref{sec:mainsec} is essentially  providing the \textit{intra-wire pairing (spin-gap) mechanism}, which in turn, under the right conditions, can be used to construct the anisotropic superconducting phase. Conventionally, this mechanism is provided by effective attractive interactions due to interactions with phonons or other degrees of freedom, and we present here a feasibly controllable way of inducing this gap. Then, manipulating additional experimental parameters, e.g., different gate-voltages, one can achieve long range coherent superconductivity in a two-dimensional system driven solely by repulsive electron-electron interactions.

\section{\label{sec:conclusions} Discussion and conclusions}
In this work we have presented a novel setup, which allows inducing a pairing instability for electrons in a spinfull 1D system in a tunable manner.
Our proposal utilizes strong repulsive interactions with a highly-correlated secondary system properly tuned to a desired state.
These so-called primary-secondary interactions tend to bind electrons of opposite spin in the primary wire, and in certain cases overcome their intrinsic repulsive interaction.

In the limit of weak interactions between the primary and secondary wires (or large Fermi momenta mismatch), the two are effectively decoupled (up to innocuous bilinear term which do not open a gap), and the primary wire is in a LL phase. In the opposite limit of strong inter-wire interactions and near commensuration, we find a phase of bound trions, involving a spin-singlet pair of primary electrons and one secondary fermion. This trion phase displays a single-particle gap in both the primary and secondary parts, and can lead to fractional transport signatures. The trions tend to have CDW order, yet in an extreme limit may also have superconducting tendencies.

Using a perturbative RG analysis, we have established the paired phase of spin-singlet primary electrons as the ground-state in an intermediate region of phase space, between the trion and decoupled phases mentioned above. Similarly to the trion phase, quasi-long-range CDW order generically has the largest susceptibility in this paired phase, whereas the superconducting tendencies are dominant in a narrow (though sensible) range of parameters.

The RG treatment we employed revealed two key aspects in the proposed setup. First, the competition between the intrinsic electron repulsion in the primary wire and the effective attraction mediated by the secondary sector becomes much more transparent. Eventually, at a later stage of the RG ``flow'', a crossover can occur where the competition becomes cooperation, signaling the pairing instability.

Secondly, the importance of strong correlations in the secondary fermion sector becomes clear. The phase space for the non-trivial (paired and trion) phases is enlarged by the secondary repulsive interactions. This is because they allow the effective attraction to effectively overcome the intrinsic (primary) repulsion, even in scenarios where the bare value of the former is somewhat smaller than the latter. (This is of course the more generic scenario, if the interaction becomes weaker with larger spatial separation.)

Our conclusions regarding the nature of the electron paired phase are not unlike those drawn from studies of previous models of repulsion-mediated attraction in 1D systems.
We have demonstrated the relation between these models and a certain limit of our proposal, and discussed the key differences between them.
Our model provides a continuum description of the pairing mechanism, which originates in (nearly) momentum conserving backscattering processes involving mobile charge carriers, instead of stationary two-level polarizers.
Moreover, the analysis presented in this work takes into account interactions between the secondary degrees of freedom (previously not considered), and reveals their significance.

Extension of our proposal to an anisotropic 2D system was shown to possibly facilitate long-range phase coherent superconductivity. Remarkably, this can be achieved in a system with only repulsive interactions. The role of our proposal is supplying the mechanism for pairing from repulsion, whereas the discussion on inter-wire phase coherence (brought on by pairs hopping between neighboring primary wires) is mostly already well-established in the literature.

An accessible method by which pairing between electrons in different bands can be engineered opens up new and exciting possibilities for condensed matter research and experiments.
The model we study, while motivated by a desire to manufacture and design effective attractive forces between electrons, may also elucidate the manner by which electrons pair in other low-dimensional strongly-correlated materials, and possibly further the pursuit of higher-$T_c$ superconductors and other unconventional superconductors, e.g., magic angle twisted-bilayer graphene~\cite{Cao2018}.

\begin{acknowledgments}
This project was partially supported by grants from the ERC under the European Union’s Horizon 2020 research and innovation programme (grant agreement LEGOTOP No 788715), the DFG CRC SFB/TRR183, the BSF and NSF (2018643), the ISF (1335/16), and the ISF Quantum Science and Technology (2074/19).

\end{acknowledgments}

\appendix

\section{CNT as a secondary system}\label{app:CNTsecondaryproposal}
In this Appendix we demonstrate that with proper gating, a combination of the intrinsic spin-orbit coupling in a CNT and magnetic flux in the direction of the nanotube axis enables one to tune the CNT to a point with a single conducting spin-polarized band. This is most conveniently shown for the case of a zig-zag nanotube, though other chiralities may also suffice.

We write the continuum low-energy (single-particle) Hamiltonian,
\begin{equation}
    H_{\rm zz}\left(k\right)=v_{F}k\rho_{y}+\left(\Delta_{0}+\Delta_{{\rm so}}\sigma_{z}\nu_{z}+\Delta_{{\rm \Phi}}\nu_{z}\right)\rho_{x}-\mu,\label{eq:H0zigzag}
\end{equation}
with $\rho_{i},\sigma_{i},\nu_{i}$ Pauli matrices operating on the sublattice, spin, and valley subspaces, respectively, $k$ is the momentum along the CNT axis, $\Delta_{{\rm so}},\Delta_{\Phi}$ are the energy gaps associated with the spin-orbit coupling and the flux, respectively \cite{IlaniSOC}, $\Delta_0$ accounts for the gap in the CNT spectrum (in the absence of spin-orbit and magnetic fields), due to either the curvature \cite{CurvatureGap} or the chirality of the nanotube, $v_F$ is the Fermi velocity, and $\mu$ is the chemical potential.

In \eqref{eq:H0zigzag} we have assumed $\Delta_0>\Delta_{\rm so}$, so that a Zeeman-like spin-orbit term is effectively absorbed into $\Delta_{\rm so}$. We note that this Zeeman spin-orbit term, as well as the curvature gap, vanish for armchair CNTs.

\begin{figure}
\includegraphics[scale=0.55]{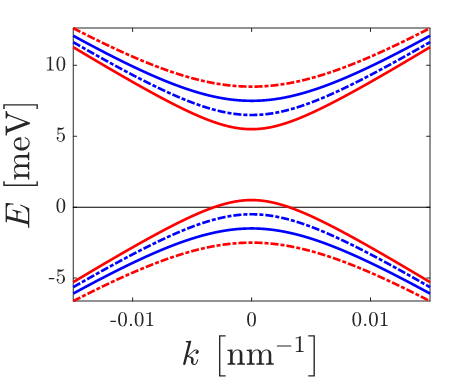}
\caption{\label{fig:auxSpectrumcnt}
Example of a CNT spectrum tuned to the single occupied band regime. The spectrum was calculated according to Eq. \eqref{eq:cntauxspectra}, with the parameters $v_F=8\cdot 10^5\,\rm{\frac{m}{sec}}$, $\Delta_0=4\,\rm{meV}$,  $\Delta_{\rm{so}}=1\,\rm{meV}$,  $\Delta_\Phi=0.5\,\rm{meV}$, and $\mu=-3\,\rm{meV}$. Different colors and line styles correspond to different spin projections and valley indices, respectively. The Fermi energy is marked by a solid black line.
}
\end{figure}

The spectrum decomposes into eight bands with well-defined spin ($\sigma=\pm1$) and valley ($\nu=\pm1$) labels,
\begin{equation}
    E_{\sigma,\nu}=-\mu\pm\sqrt{\left(v_{F}k\right)^{2}+\left(\nu\Delta_{0}+\sigma\Delta_{{\rm so}}+\Delta_{{\rm \Phi}}\right)^{2}}.\label{eq:cntauxspectra}
\end{equation}
As an example, Fig. \ref{fig:auxSpectrumcnt} shows an example where the CNT is tuned such that the Fermi energy crosses a single spin and valley polarized hole-like band.

\section{Role of the mass in the secondary sector}\label{app:massRG}
The possibility of the secondary sector having a bare gap was mostly discussed in Sec. \ref{sec:Littlemodel} of the main text, where the connection to the original setup proposed by Little was discussed. Clearly if the mass term completely overwhelms the energy scales associated with the inter-wire interactions, $V_\phi,g_{1,2}$, we may effectively set ${\cal H}_{\rm int}\to 0$, and we get none of the effects described in this work. More accurately, one may integrate out the auxiliary fermions completely, finding a correction to $K_c$ due to a term $\propto\frac{V_\phi^2}{m_{\rm sec}}\left(\partial_x \phi_c\right)$ generated.

However, when the mass term competes with ${\cal H}_{\rm int}$, one may still recover most of the interesting physics we have uncovered. This may be understood by considering the simplest possible way $m_{\rm sec}$ affects the RG flow: an additional length scale $L_m\approx\frac{u}{\left|m_{\rm sec}-\mu_{\rm sec}\right|}$, above which the system ``realizes'' that the secondary fermions are in fact gapped (and therefore cannot induce backscattering), and $g_{1,2}$ get cut-off. The appropriate RG-time cutoff for the mass term is
\begin{equation}
    \ell^{m,*}=\ln\frac{\Lambda_{\rm sec}}{\left|m_{\rm sec}-\mu_{\rm sec}\right|},\label{eq:massCutoff}
\end{equation}
with the effective secondary bandwidth $\Lambda_{\rm sec}\equiv\frac{u_{\rm sec}}{\alpha_0}$. Thus, the presence of a sufficiently small (or tunable) gap for the secondary fermions provides an additional degree of freedom to tune between the phases of the system.

We note that this simplified treatment of the mass term, much like the way we handled the incommensurability, is only approximate, as $m_{\rm sec}$ flows itself, and affects some of the other coupling coefficients. However, the qualitative picture, in terms of how the gap impacts the low-energy behavior of the system, should remain intact.

\section{The trion phase}\label{app:lockeddetails}
We begin by omitting the spin degree of freedom, assuming it has been gapped out, and absorbing $\left\langle \cos\left(\sqrt{2}\phi_{s}\right)\right\rangle $ into the definition of the relevant $g$ term, to get a modified $g^*$ coupling. Manipulating the Hamiltonian given by Eqs. \eqref{eq:Htotal}, \eqref{eq:Hc}, \eqref{eq:Habososn}--\eqref{eq:Hintboson}, and writing in in terms of the sectors defined in \eqref{eq:lockedsectors}, we may write the effective Hamiltonian as \begin{subequations}
\begin{equation}
    {\cal H}_{{\rm lock}}={\cal H}_{f}+{\cal H}_{g}+{\cal H}_{\times},
\end{equation}
\begin{equation}
    {\cal H}_{f}=\frac{u_{f}}{2\pi}\left[\frac{1}{K_{f}}\left(\partial_{x}\phi_{f}\right)^{2}+K_{f}\left(\partial_{x}\theta_{f}\right)^{2}\right],
\end{equation}
\begin{align}
        {\cal H}_{g}&=\frac{u_{g}}{2\pi}\left[\frac{1}{K_{g}}\left(\partial_{x}\phi_{g}\right)^{2}+K_{g}\left(\partial_{x}\theta_{g}\right)^{2}\right]\nonumber\\
        &+\frac{g^{*}}{2\pi^{2}a^{2}}\cos\left(\sqrt{6}\phi_{g}+2\left(k_{F}+k_{a}\right)x\right),
\end{align}
\begin{equation}
    {\cal H}_{\times}=\frac{V_{\times}}{\pi^{2}}\partial_{x}\phi_{g}\partial_{x}\phi_{f}+\frac{V'_{\times}}{\pi^{2}}\partial_{x}\theta_{g}\partial_{x}\theta_{f}.
\end{equation}
\end{subequations}
Apart from the expression for $K_f$, mentioned in the main text, the other coefficients going into this Hamiltonian are given by \begin{subequations}\label{eq:appBparameters}
\begin{equation}
    K_g=\sqrt{\frac{u_{c}K_{c}+2u K+\sqrt{8}\frac{V_{\theta}}{\pi}}{\frac{u_{c}}{K_{c}}+\frac{2u}{K}+\sqrt{8}\frac{V_{\phi}}{\pi}}},
\end{equation}
\begin{equation}
    u_{f}=\frac{1}{3}\sqrt{\left[ \frac{2u_{c}}{K_{c}}\frac{u}{K}-\sqrt{8}\frac{V_{\phi}}{\pi}\right] \left[ 2u_{c}K_{c}+u K-\sqrt{8}\frac{V_{\theta}}{\pi}\right] },
\end{equation}
\begin{equation}
    u_{g}=\frac{1}{3}\sqrt{\left[ \frac{u_{c}}{K_{c}}+\frac{2u}{K}+\sqrt{8}\frac{V_{\phi}}{\pi}\right]\left[ u_{c}K_{c}+2u K+\sqrt{8}\frac{V_{\theta}}{\pi}\right] },
\end{equation}
\begin{equation}
    V_{\times}=\frac{1}{3}\left[\sqrt{2}\pi\left(\frac{u_{c}}{K_{c}}-\frac{u}{K}\right)+V_{\phi}\right],
\end{equation}
\begin{equation}
    V'_{\times}=\frac{1}{3}\left[\sqrt{2}\pi\left(u_{c}K_{c}- u K\right)+V_{\theta}\right].
\end{equation}
\end{subequations}
The Hamiltonian ${\cal H}_{{\rm lock}}$ can be used as a starting point to more accurately capture the commensurate-incommensurate transition the system goes through when transitioning from the (spin-gapped) paired phase to the trion one.

Expanding the trion operator $\Psi_{\rm trion}$ in terms of the bosonic variables, one finds
\begin{widetext}
\begin{align}
    \Psi_{{\rm trion}}&=\frac{1}{\sqrt{2}}\left(\frac{1}{\pi \alpha}\right)^{\frac{3}{2}} e^{i\sqrt{3}\theta_{f}}\left[\cos\left(\sqrt{\frac{8}{3}}\phi_{g}+\sqrt{\frac{1}{3}}\phi_{f}-\left(2k_{F}+k_{\rm sec}\right)x\right)+\cos\left(\sqrt{3}\phi_{f}-\left(2k_{F}-k_{\rm sec}\right)x\right)\right]\nonumber\\
    +&\frac{1}{\sqrt{2}}\left(\frac{1}{\pi \alpha}\right)^{\frac{3}{2}}e^{i\sqrt{3}\theta_{f}}\left[\cos\left(\sqrt{2}\phi_{s}\right)\right]\cos\left(\sqrt{\frac{2}{3}}\phi_{g}-\sqrt{\frac{1}{3}}\phi_{f}-k_{\rm sec}x\right).
\end{align}
\end{widetext}
In the trion phase, taking also $\left|k_{\rm sec}\right|=\left|k_F\right|$, one recovers the most relevant contribution, Eq. \eqref{eq:trionbosonized} in the main text.

Regarding backscattering impurity operators, one can distinguish three possible kinds: (i) Impurities in the electronic system, which backscatter both spins equally, of the bosonic form $~\cos\left(\sqrt{2}\phi_{s}\right)\cos\left(\sqrt{2}\phi_{c}\right)$; (ii) Impurities which impact a single spin channel $\sigma$, $~\cos\left(\sqrt{2}\left(\phi_{c}+\sigma\phi_{s}\right)\right)$; (iii) Impurities in the secondary wire, $~\cos\left(2\phi\right)$. Deep in the trion phase, all three are proportional to $\cos\left(\frac{2\phi_{f}}{\sqrt{3}}\right)$. This makes the impurities irrelevant in the RG sense once $K_f>3$.

\section{Fractional conductance}\label{app:fractionalconductancecalc}
We briefly give here the derivation for the fractional conductance, along the lines described in Ref.~\cite{fractionalGSYO} and its Supplementary Materials.
We consider a setup where non-interacting leads are adiabatically attached to both the primary and secondary wires, and consider the scattering problem of incoming and outgoing currents in this system. These currents are related by
\begin{equation}
    \begin{pmatrix}O_{R}\\
O_{L}
\end{pmatrix}=\begin{pmatrix}\mathcal{T} & 1-\mathcal{T}\\
1-\mathcal{T} & \mathcal{T}
\end{pmatrix}\begin{pmatrix}I_{R}\\
I_{L}
\end{pmatrix},\label{eq:Tmatrix}
\end{equation}
where $O_{R.L}$ and $I_{R,L}$ are chiral outgoing and incoming current vectors of length $3$, corresponding to the total number of modes in the system: primary spin-up (enumerated $i=1$), primary spin-down ($i=2$), and the secondary spinless mode ($i=3$), whereas $\mathcal{T}$ is a $3\times 3$ matrix. In terms of the $\phi_{i}$ bosonic variables, the current elements are
\begin{equation}
    I_{R,i} = \frac{e}{2\pi}\partial_t\frac{\theta_i-\phi_i}{\sqrt{2}}|_{x=\frac{L}{2}},\,\,\,\, I_{L,i} = \frac{e}{2\pi}\partial_t\frac{\theta_i+\phi_i}{\sqrt{2}}|_{x=-\frac{L}{2}},
\end{equation}
\begin{equation}
    O_{R,i} = \frac{e}{2\pi}\partial_t\frac{\theta_i-\phi_i}{\sqrt{2}}|_{x=-\frac{L}{2}},\,\,\,\, O_{L,i} = \frac{e}{2\pi}\partial_t\frac{\theta_i+\phi_i}{\sqrt{2}}|_{x=\frac{L}{2}}.
\end{equation}
We move to a new basis spanning this 3D space, with vectors $\mathbf{n}_s=\frac{1}{\sqrt{2}}\left(1,-1,0\right)^T$, $\mathbf{n}_g=\frac{1}{\sqrt{6}}\left(1,1,2\right)^T$, and $\mathbf{n}_f=\frac{1}{\sqrt{3}}\left(1,1,-1\right)^T$, corresponding to the spin, gapped, and ``free'' (LL) sectors, respectively. Notice that these three vectors form an orthonormal set.

Deep in the so-called trion phase, the fields $\phi_s,\phi_g$ are pinned throughout the system, enforcing the boundary conditions
\[
\partial_t \phi_\uparrow-\partial_t \phi_\downarrow=0,
\]
\[
\partial_t \phi_\uparrow+\partial_t \phi_\downarrow+2\partial_t \phi=0.
\]
Taken at opposite ends of the system, this boundary condition is leads to
\begin{equation}\label{eq:fraccond1}
    \mathbf{n}_s^T\mathcal{T}=0,\,\,\,\,\,\mathbf{n}_g^T\mathcal{T}=0.
\end{equation}
The unobstructed propagation of the $\phi_f$ mode through the system leads to the boundary conditions \[O_{R/L,1}+O_{R/L,2}-O_{R/L,3}=I_{R/L,1}+I_{R/L,2}-I_{R/L,3},\] or equivalently,
\begin{equation}
    \mathbf{n}_f^T\mathcal{T}=n_f^T. \label{eq:fraccond2}
\end{equation}
The solution to Eqs. \eqref{eq:fraccond1} and \eqref{eq:fraccond2} can be readily found to be \[\mathcal{T}=1-\mathbf{n}_s\mathbf{n}_s^T-\mathbf{n}_g\mathbf{n}_g^T.\]

The total current flowing through the primary wire may be expressed as $J=\left(1,1,0\right)\cdot\left(I_R-O_l\right)$. Assuming that in the primary incoming right movers emanate from a reservoir at potential $V$ and the left movers from a reservoir with zero potential, and that the secondary wire is grounded,we set $I_R=\frac{e^2}{h}V\left(1,1,0\right)^T$, and $I_L=\left(0,0,0\right)^T$.
The two-terminal conductance \textit{measured at the primary wire terminals} can then be extracted,
\begin{equation}
    \frac{G}{{e^2}/{h}}=\left(1,1,0\right)\mathcal{T}\left(1,1,0\right)^T=\frac{4}{3}.
\end{equation}
The current flowing in the secondary wire in this setup is simply $\left(0,0,1\right)\cdot\left(I_R-O_l\right)$, which is precisely $-\frac{2}{3}\frac{e^2}{h} V$.
Other transport coefficients may also be calculated using the same $\mathcal{T}$ matrix obtained here.

We briefly comment on the setup where the secondary wire is disconnected from any leads, and is ``floating''. In the trion phase, where the charge in the primary wire is ``locked'' to the secondary holes, we expect zero current to flow in the primary wire (as long as the gap in the $\phi_g$ sector is large enough). This is reminiscent of the absolute drag between two capacitively coupled quantum wires~\cite{SternColumbDrag}, and is essentially a generalization of this phenomenon.

\bibliography{Draft}

\end{document}